\documentclass[pra,aps,twocolumn]{revtex4}
\usepackage{graphicx}
\usepackage{color}
\usepackage{epsfig}

\begin{document}
\title{Light scattering in inhomogeneous Tomonaga-Luttinger liquids}

\author{E. Orignac}
\affiliation{Laboratoire de Physique de l'\'Ecole Normale
  Sup\'erieure de Lyon, CNRS-UMR5672, 69364 Lyon Cedex 7, France}

\author{R. Citro}
\affiliation{Dipartimento di Fisica "E. R. Caianiello" and Spin-CNR,
Universit\`a degli Studi di Salerno, Salerno, Italy}

\author{S. De Palo}
\affiliation{IOM-CNR and Dipartimento di Fisica Teorica, Universit\`a
  Trieste, Trieste, Italy }

\author{M.-L. Chiofalo}
\affiliation{INFN and Dpt. of Mathematics,
  University of Pisa, Pisa, Italy}

\begin{abstract}

   We derive the dynamical structure factor for an
  inhomogeneous Tomonaga-Luttinger liquid as can be formed  in a confined
  strongly interacting one-dimensional gas. In view of current
  experimental progress in the field, we provide
  a simple analytic expression for the light-scattering cross
  section, requiring only the knowledge of the density dependence of the
  ground-state energy as they can be extracted {\sl e.g.} from exact or
  Quantum Monte Carlo techniques, and a Thomas-Fermi
  description. We apply the result to the case of
  one-dimensional quantum bosonic gases with dipolar interaction in a
  harmonic trap, using an energy functional deduced from  Quantum Monte Carlo
  computations. We find an universal
  scaling behavior peculiar of the Tomonaga-Luttinger liquid, a signature that
  can be eventually probed by Bragg spectroscopy in
  experimental realizations of such systems.
\end{abstract}

\pacs{71.10.Pm,67.85.-d,67.10.Hk}

\maketitle

\section{Introduction}

It is well known theoretically
that systems of reduced dimensionality, especially in one
dimension, present simultaneously enhanced
quantum fluctuations and stronger interaction effects that can lead to
exotic ground
states~\cite{varma_nonfermiliquid_review,schulz_houches_revue}.
>From the experimental point of view, there
are many prototypical one-dimensional systems, that range from
organic~\cite{bourbonnais_jerome_review,bourbonnais2008} or
inorganic~\cite{mizokawa2002,wang06_arpes_limo6o17} conductors,
and antiferromagnetic (AF) spin chain~\cite{hammar1999,lake2005} or
ladder~\cite{dagotto_supra_ladder_review,klanjsek08_bpcp} materials,
to  nanoscale systems such as
of quantum wires~\cite{auslaender_quantumwire_tunneling,hilke2001},
carbon
nanotubes~\cite{bockrath_luttinger_nanotubes,ishii03_nanotube_pes,gao2004,lee2004}
or self organized Au atomic wires on Ge(001) semiconductor
surfaces~\cite{blumenstein2011}. More
recently, advances in atom trapping technology has permitted the
realization of both fermionic and  bosonic one-dimensional systems with
unprecedented control~\cite{paredes_toks_experiment,kinoshita_tonks_experiment,liao2010,cazalilla_review_11}.  The low-energy physics of
such one-dimensional systems is well described by the Tomonaga-Luttinger Liquid (TLL) theory
\cite{schulz_houches_revue,varma_nonfermiliquid_review,voit_bosonization_revue,giamarchi_book_1d}. In
a single component  TLL, there is a single gapless branch of
excitations with linear dispersion,
and the  interplay between interactions and quantum fluctuations in
the ground state leads to
power-law decay of correlations with interaction dependent
exponents. Remarkably, the low energy theory is fully characterized by
two parameters: the velocity $u$ of the linearly dispersion
excitations, and the dimensionless exponent $K$
controlling the decay of all correlations, the corresponding exponents
being rational functions of $K$.  In physical systems several prominent
features of TLL have been observed after measuring
the spectral
function~\cite{bourbonnais_jerome_review,wang06_arpes_limo6o17}, the
structure factor~\cite{lake2005} or the conductivity~\cite{bockrath_luttinger_nanotubes}, and more recently the first
quantitative check of TLL physics has appeared for the spin-1/2
ladder material bis(piperidinium)
tetrabromocuprate(II) (C$_5$H$_{12}$N)$_2$CuBr$_4$  (abbreviated BPCB in the following), in an applied magnetic field~\cite{klanjsek08_bpcp}.
However, despite this recent achievement, in many of the physical systems mentioned above, little control can be exerted on the values of $u$ and
$K$ and thus the Luttinger exponent $K$  is taken as an
adjustable parameter~\cite{bourbonnais_jerome_review,bockrath_luttinger_nanotubes}. This fact prompts for the search of more than
one signature of Tomonaga-Luttinger liquid physics for a single
system.

In the case of systems with strong
confinement (e.g. confined quantum gases), excitation properties
can be most easily accessed by light spectroscopy techniques, as
proposed in the early days of atomic Bose-Einstein
Condensation~\cite{Javanainen95,Graham96}. For
example, the spectral function has recently been measured in trapped
Fermi gases by radiofrequency spectroscopy~\cite{one} and the dynamical
structure factor has been studied successfully by optical Bragg
spectroscopy in free and trapped Bose-Einstein condensates~\cite{two,three, four, five, six} as well as trapped Fermi gases~\cite{seven}.
Bragg spectroscopy can be based on energy transfer to the system at
fixed momenta~\cite{eleven,twelve,thirth} or can permit the study of
the full momentum composition of excitations by a coherent momentum
transfer mapping~\cite{ernst_bragg}. For these reasons,
Bragg spectroscopy can be especially useful to investigate
the properties of the many phases
realizable in these systems such as Mott insulator, Tonks-Girardeau
gas or supersolid phases as recently
proposed~\cite{fourth,fiveth,sixth,seventh,eightth,nineth,twentyth,
twentyoneth,golovach_2009}. The most recent experimental progress in producing
long-lived ground-state polar molecules in
a three-dimensional (3D) optical lattice, and possibly also in 1D arrays of
pancakes and 2D arrays of tubes~\cite{Chotia_2011} as well as
condensates of dipolar atoms~\cite{pfau_a0,lu2011}, opens up wide perspectives
in the comprehension of controlled quantum systems with tunable short and long
range interactions under progressively reduced dimensionality.

As we have more extensively reviewed in Ref.~\cite{ours}, among many
possible realizations, quantum dipolar
gases in 1D confinement are quite peculiar TLL systems. Here in fact, one single
parameter drives the crossover from weak to strong interaction
regimes, where however the weakest regime is a
Tonks-Girardeau state, the strongest being a Density Dipolar Wave
state
characterized by quasi-ordering.
Based on the above motivations, we derive an
analytic expression for light-scattering intensity in
the case of a weakly inhomogeneous TLL. This expression is valid within a
Thomas-Fermi description, where the system can be considered locally
homogeneous. The expression requires the knowledge of the density
dependence of the  ground-state energy of the homogeneous system, as
 can be obtained by {\em e.g.} approximate calculations, exact
Bethe-Ansatz technique or
Quantum Monte Carlo (QMC) simulations. The paper is organized as follows.
After reviewing in
Sec.~\ref{sec:LS_h} the calculation of the dynamic structure
factor and the inelastic light-scattering cross section of homogeneous
Tomonaga-Luttinger Liquids, we derive in Sec.~\ref{sec:LSin_h} the general
expression for the inhomogeneous system within the Thomas-Fermi
approach, in terms of the eigenvalues and
eigenfunctions of the hydrodynamic TLL.
We then specialize in Sec.~\ref{sec:TLLdip} to the case of
one-dimensional quantum bosonic gases with dipolar interaction in an
harmonic trap, using our previous QMC
findings~\cite{ours}. Here the results are explicitly discussed
in the various regime while the single parameter built-up from
density and interaction strength is tuned.

\section{Light-scattering cross section in homogeneous
Tomonaga-Luttinger Liquids }~\label{sec:LS_h}

 The dynamic structure factor $S(\boldmath{q},\omega)$ is central
  in the description of  interacting many-body  systems.
  $S(\boldmath{q},\omega)$ is related to the Fourier transform of the
  imaginary  density-density
  correlation function with the fluctuation-dissipation theorem.
  It is therefore accessible
  by means of inelastic scattering, where density fluctuations are
  induced in the system and their subsequent relaxation
  is measured revealing the system characteristics.
While inelastic neutron
  scattering has been the tool to probe the condensate nature of
  superfluid helium and the roton spectrum~\cite{Sokol_BEC},
  inelastic light scattering has been proposed
  and widely used in dilute quantum degenerate
  gases. Within linear response theory the scattering
  cross section $\sigma$ of light at frequency $\omega$ and
  angle $\Omega$ incident on a Bose atomic sample is:
\begin{equation}
  \label{eq:cross-section}
  \frac{d^2\sigma}{d\Omega d\omega} \propto \frac{1}{\pi
    n}(n_B(\omega)+1) \mathrm{Im} \chi(q,\omega)=S(q,\omega)\; ,
\end{equation}
where $n_B(\omega)$ is the Bose distribution function, $n=N/V$ and
$\chi(q,\omega)$  is the Fourier transform of the density-density
correlation function
\begin{equation}
\chi(r,t)=-i \theta(t)\langle[n(r,t),n(0,0)]\rangle.
\end{equation}

Earlier experimental
studies~\cite{three} have shown that condensate properties of atomic cold gases
could be studied  by means of Bragg scattering yielding large energy
resolution and sensitivity.  The system is illuminated by two lasers
beams of momenta $\boldmath{k}_1$ and $\boldmath{k}_2$ and frequencies $\omega_1, \omega_2$ of difference $\omega$ that creates a periodic field whose intensity is
proportional to $\cos{[(\boldmath{k}_1-\boldmath{k}_2)\cdot
    \boldmath{r}-\omega t]}$. The external potential couples to the
density $ n(\boldmath{q})$ of the system where
$\boldmath{q}=\boldmath{k}_1-\boldmath{k}_2$. After using the golden
rule, the response of the
system to this perturbation is the  dynamical
structure factor~\cite{brunello2001}.   Light scattering experiments then
directly measure $S(\boldmath{q},\omega)$.

This quantity is then a benchmark against the theoretical descriptions
of the systems.  For an homogeneous Tomonaga-Luttinger liquid occuring in
interacting one-dimensional system the dynamic  structure factor can
be readily obtained~\cite{giamarchi_book_1d}. In the following,
we briefly sketch the derivation.
For a system of interacting spinless particles, either bosons or fermions, the low-energy physics is that of a
Tomonaga-Luttinger liquid whose Hamiltonian  is
\begin{equation}
  \label{eq:hamiltonian-gases}
  H=\int \frac{dx}{2\pi} \left[ u K (\pi \Pi)^2 + \frac u K
    (\partial_x \phi)^2 \right],
\end{equation}
with $u$ the velocity of the excitations and $K$ the Tomonaga-Luttinger
exponent. The density operator $n(x)$ is
expressed in terms of bosonic operators $\phi$:
\begin{equation}
  \label{eq:bosonized-cold}
  n(x)=n_0 -\frac 1 \pi \partial_x \phi + \sum_m A_m \cos (2m(
  \phi(x) - \pi n_0 x)),
\end{equation}
with $m$ an integer and $n_0$ the equilibrium density.

If the wavelength of the incoming light is
much larger than the average interparticle distance, we can neglect
the contribution of the oscillatory terms in
Eq.~(\ref{eq:bosonized-cold}).  Using translational
invariance, the expression for the density-density  response function becomes:
\begin{equation}
\label{eq:boso-commutator}
\chi(x-x',t)=  i \frac{\theta(t)}{\pi^2}  \langle
    [\partial_x \phi(x,t), \partial_x \phi(x',0)]\rangle.
\end{equation}
Knowing that the time-ordered correlation function $\langle T_\tau \lbrack
\phi(x,\tau)-\phi(0,0)\rbrack^2\rangle=K F_1(x,\tau)$ with
$F_1(x,\tau)=
\log \lbrack {(x^2+(u|\tau|+a)^2)/a^2}\rbrack/2$, the imaginary part
of the response function (\ref{eq:boso-commutator})
can be obtained~\cite{giamarchi_book_1d} as
\begin{equation}
\label{eq:ima_chi_ho}
\mathrm{Im} \chi(q,\omega)=\frac{q^2}{2 \omega} u K \left[
  \delta(\omega+ u |q|)-\delta(\omega-u|q|)\right]\; \; ,
\end{equation}
giving the scattered intensity at zero temperature:
\begin{eqnarray}
  \label{eq:intens-hom}
  \frac{d^2\sigma}{d\Omega d\omega} \propto S(q,\omega)&=&
  \mathrm{sign}(\omega)\mathrm{Im} \chi(q,\omega)\nonumber \\
 &=& \frac{K |q|} 2 [\delta(\omega +
  u |q|)+ \delta(\omega - u |q|)].
\end{eqnarray}
Expression (\ref{eq:intens-hom})
embodies the symmetry with respect to inversion of the velocity $u$ as
required by Galilean invariance, and evidences the dependence
of the light-scattering signal from the ratio $q/\omega$.

\section{Light-scattering cross section in inhomogeneous
Tomonaga-Luttinger Liquids}~\label{sec:LSin_h}
\subsection{Hydrodynamic approach}\label{sec:hydro}
The presence of an external potential $V(x)$ confining the
cold atomic cloud induces density inhomogeneity, and the external
light perturbation probing the density-density correlation function
introduces time-dependent processes.
The treatment of the problem is
easier under conditions of weak inhomogeneity and slow processes
as they can be met in experiments, where
external potentials vary on length
and time scales longer than the characteristic system
quantities, and local equilibrium hydrodynamic behavior sets in.
Under these conditions,
the gas can be still described by a hydrodynamic Tomonaga-Luttinger Liquid
Hamiltonian~\cite{maslov_pure_wire,safi_pure_wire,fazio_thermal_1d,petrov_trapped_bosons,petrov04_bec_review,ours}
\begin{equation}
\label{eq:ham-non-uniform}
H_{TLL}=\int^{R}_{-R} \frac{dx}{2 \pi} \left[ u(x)K(x)\pi^2 \Pi(x)^2
  +\frac{u(x)}{K(x)} (\partial_x \phi(x))^2 \right]\; .
\end{equation}
Here, the boundary conditions imposed are $\phi(-R)=0$ and
$\phi(R)=-\pi N$, with $N$ the number of particles in the system.
The parameters $u(x)$ and $K(x)$ now depend on position. In analogy with the
homogeneous case, where $u$ and $K$ are related by the expressions
$u/K=(\hbar\pi)^{-1}\partial\mu/\partial n$  and by
Galilean invariance $uK=\pi\hbar n/m$, one sets:
\begin{eqnarray}
\label{eq:uK}
u(x)K(x) &=& \pi \frac{\hbar}{m} n_0(x) \\ \frac{u(x)}{K(x)} &=&
\frac{1}{\hbar \pi} \left( \frac{\partial \mu(n)}{\partial
  n}\right)_{n=n_0(x)}
\end{eqnarray}
Once an estimate of the equilibrium density $n_0(x)$
and of the chemical potential $\mu(n)$ are known, this
phenomenological approach allows the determination of $u(x)$ and
$K(x)$.

The response function~(\ref{eq:boso-commutator})
in the case of the Hamiltonian~(\ref{eq:ham-non-uniform}) can be
calculated using the
decomposition:
\begin{eqnarray}
  \label{eq:decomposition-eigen}
\nonumber \phi(x)&=& -\pi \frac{\int_{-R}^x  dx' \frac
  {K(x')}{u(x')}}{\int_{-R}^R  dx' \frac{K(x')} {u(x')}} N \\ &+&
\sum_n \sqrt{\frac{\pi}{2\omega_n}} (a^\dagger_n + a_n) \varphi_n(x)\; .
\end{eqnarray}
Here, $[a_n,a^\dagger_n]=\delta_{n,m}$ and the first term comes from the addition of $N$ particles in the system. The functions $\varphi_n$ satisfy the eigenvalue equation:
\begin{eqnarray} \label{eigen}
  -\omega_n^2 \varphi_n = u(x) K(x) \partial_x \left(\frac{u(x)}{K(x)}
  \partial_x \varphi_n \right)\; ,
\end{eqnarray}
with boundary conditions $\varphi_n (\pm R)=0$, 
and the normalization
\begin{eqnarray}
\label{eq:orthogonality}
 \int dx \frac{\varphi_n(x) \varphi_m(x)}{u(x) K(x)}= \delta_{n,m}\; .
\end{eqnarray}
The influence of the trapping potential enters eq.~(\ref{eigen})
{\it via} the equations for $u(x)$ and $K(x)$ (\ref{eq:uK}).
The density-density response function thus can be expressed as:
\begin{equation}
 \label{eq:eigen-expansion}
 \chi(x,x',t)=\theta(t)\sum_n \frac{1}{\pi\omega_n}
 \frac{d\varphi_n}{dx} \frac{d\varphi_n}{dx'} \sin(\omega_n t),
\end{equation}

Taking the Fourier transforms with respect to $x$ and $x'$ and the
Laplace transform with respect to $t$, we find:
\begin{eqnarray}
  \label{eq:laplace-fourier}
 \chi(q,z)=\sum_n \frac{q^2 |\hat{\varphi}_n(q)|^2}{2\pi \omega_n}
 \left(\frac{1}{z+\omega_n} -\frac{1}{z-\omega_n}\right),
\end{eqnarray}
where $\mathrm{Im}(z)>0$.
Finally, taking the limit $z \to \omega +i0_+$ we obtain:
\begin{eqnarray}
\nonumber \mathrm{Im} \chi (q,\omega +i0_+)&=& \frac{q^2}{2\omega}
\sum_n |\hat{\varphi}_n(q)|^2 [\delta(\omega-\omega_n) \\ &+&
  \delta(\omega+\omega_n)]\; .
\label{eq:ima_chi_noho}
\end{eqnarray}
Eq.~(\ref{eq:ima_chi_noho}) maintains the structure of its homogeneous
counterpart  (\ref{eq:ima_chi_ho}).

The density-density response
function can be determined whenever the density dependence of
the ground state energy per unit length $e(n)$ or of the chemical potential
$\mu(n)=(\frac{\partial e}{\partial n})|_{n=n(x)}$
 is known. An especially simple  situation is realized  when
$e(n)\propto n^{\gamma +2}$. That type of dependence of energy
on density corresponds to several limiting cases of 1D TLL systems.
For example, in the Lieb-Liniger gas~\cite{lieb_bosons_1D,lieb_excit}
there are two well understood limits. At low density or strong
repulsion, the gas behaves as a hard-core boson
gas~\cite{girardeau_bosons1d}  with $\gamma=1$, while at high density or
weak repulsion, the Bogoliubov approximation applies and gives and
energy density proportional to  $n^2$, so that $\gamma=0$.
The study of the crossover between these two limits requires the
Bethe-Ansatz computation of the ground state energy
density~\cite{lieb_bosons_1D}. A similar situation
occurs in the case of dipolar gases. For low densities, the energy per
unit length $e(n)$ has the $\gamma=1$ behavior typical of hard core bosons,
while for high density it has the $\gamma=2$ behavior of a
crystal of classical dipoles, and a Dipolar-Density-Wave
manifests~\cite{ours}. As density increases,
the system crosses over from the low density
hardcore boson gas  to the high density Dipolar-Density-Wave.

In the model with $e(n)=g n^{\gamma+2}$ and in the case of harmonic
trapping potential $V(x)=m\Omega_0^2x^2/2$, the
eigenvalues $\omega_n$ of (\ref{eq:decomposition-eigen}) can be found
exactly, and the functions $\varphi_n$ are expressible in terms of
Gegenbauer polynomials~\cite{menotti02_bose_hydro1d,petrov04_bec_review} as:
\begin{eqnarray}
\label{eq:ultraspherical}
  \varphi_n(x) &=& A_n \left(1-\frac {x^2}{R^2}\right)^{\alpha+1/2}
  C_n^{(\alpha+1)}\left(\frac x R\right),  \\ \omega_n^2 &=& \frac
  {u_0^2}{R^2} (n+1) (n+2\alpha+1)\; .
\end{eqnarray}
Here, $u_0$ and $K_0$ are the Tomonaga-Luttinger parameters corresponding to
the density at the trap center,
\begin{equation}
A_n=\sqrt{\frac{u_0 K_0}{R}
    \frac {n! (n+\alpha+1)}{\pi \Gamma(n+2\alpha+2) }} 2^{\alpha+1/2}
  \Gamma(1+\alpha)\; ,
\end{equation}
and $\alpha=(\gamma+1)^{-1} - 1/2$. In particular,
in the case of hard-core Bose gas when $\gamma=1$, $\alpha=0$ and
the Gegenbauer polynomials reduce to Chebyshev
polynomials~\cite{abramowitz_math_functions}.  In order to calculate
the scattered light intensity, we need the Fourier transform of the
$\varphi_n$'s.  Using Eq.~(7.321) of
Ref.~\onlinecite{gradshteyn80_tables} we obtain:
\begin{eqnarray}
  \label{eq:integral-gegenbauer}
   |\hat{\varphi}_n(q)|^2 &=& 2 u_0 K_0 R (n+\alpha+1)
   \frac{\Gamma(n+2\alpha+2)}{\Gamma(n+1)}\nonumber\\
    && \times \frac{J_{n+\alpha+1}^2(qR)}{(qR)^{2\alpha+2}}\; .
\end{eqnarray}
where the $J_m$ are the Bessel functions of the first kind.
Thus:
\begin{eqnarray} \label{eq:final-boso}
   &&\mathrm{Im} \chi(q,\omega+i0_+)= \frac{u_0 K_0}{R \omega} \sum_n
   (n+\alpha+1)  \frac{\Gamma(n+2\alpha+2)}{\Gamma(n+1)}\nonumber\\
  && \times \frac{J_{n+\alpha+1}^2(qR)}{(qR)^{2\alpha}}
        [\delta(\omega-\omega_n)+ \delta(\omega+\omega_n)]
\end{eqnarray}

Eq. (\ref{eq:final-boso}) shows the main features of the scattered
light intensity. This is a set of discrete peaks, whose weight is a function
of $qR$, and whose spacing reduces with increasing the trap size
$R\to \infty$.

\subsection{Approach {\it via}
Density-Functional Theory with Local Density Approximation}
\label{sec:approx}

In the present section we derive an approximate expression for the
dynamical structure factor of an inhomogeneous 1D TLL,
reverting to the Density Functional Theory (DFT) accompanied by a Local Density
Approximation (LDA). We sketch in the following the main concepts and
derivation.
Through the Hohenberg and Kohn theorem, DFT
establishes that
the ground state energy of a system subjected to an external potential
$V(x)$ is a functional
$E_g[n(x)]=E[n(x)]+ \int_{-\infty}^{+\infty} n(x) V(x)dx$
of the density $n(x)$, where $E[n(x)]$
embodies the kinetic and exchange-correlation parts.
The equilibrium density profile is determined by the variational
condition
\begin{eqnarray}
\label{eq:DFT}
\frac{\delta E_g[n(x)]}{\delta n(x)}=\mu\; ,
\end{eqnarray}
stating that equilibrium corresponds to a minimum of the energy against
changes in the particle density, while the total number of particles
is fixed through the (density-dependent)
chemical potential $\mu$. Eq. (\ref{eq:DFT})
reminds the Thomas-Fermi equilibrium condition in non-interacting
systems, and in fact the Density Functional sets a one-to-one
correspondence between the ground state energies of an interacting
system and of its non-interacting analogue.
Whenever an analytic expression of $\mu(n)$ is available,
inversion of the equation of state (\ref{eq:DFT}) allows the
determination of the equilibrium density $n_0(x)$.

While Eq. (\ref{eq:DFT}) is exact, the actual determination of
the functional $E[n(x)]$ needs approximations. Under the conditions of
shallow confinement,
we can safely use the Local Density Approximation. Here,
the functional $E[n(x)]$ is replaced by
\begin{equation}
\label{eq:LDAfunc}
E^{LDA}[n(x)]=\int
e^{hom}[n(x)] n(x) dx \; ,
\end{equation}
where $e^{hom}(n)$ is the energy per particle of the
homogeneous system with density $n$.

Differentiating $E^{LDA}[n(x)]=E_g[n(x)]+\int dx (V(x)-\lambda) 
n(x)$ with respect to $n(x)$, $\lambda$ being a Lagrange multiplier 
fixing the total number of particles, one obtains the condition for
the  local chemical potential
\begin{equation}
\label{eq:thomas-fermi_1}
 \mu[ n(x)]= V(R)-V(x),
\end{equation}
where the local chemical
potential is defined by the functional derivative:
\begin{eqnarray}
\label{eq:equation_of_state}
  \mu(n)=\frac{\delta E}{\delta  n(x)}=\left(\frac{\partial (n e^{hom}(n))}
{\partial n}\right)_{n=n_0(x)}\; .
\end{eqnarray}
If an analytic expression of $\mu(n)$ is given,
Eq.(\ref{eq:equation_of_state}) would allow to find $ n(x)$ by
inverting the relation Eq.~(\ref{eq:thomas-fermi_1}).
The energy $e^{hom}(n)$ can be
obtained after perturbation theory, or by exact calculations such as
Bethe-Ansatz, or else by computational Quantum Monte Carlo methods.

We now turn to the problem of determining the dynamical structure factor of the
inhomogeneous system. To this aim,
we follow the reasoning in ~\cite{vignolo2001,golovach_2009} and
imagine to slice it into small segments of length $\Delta x$, where the
density $n_0(x)$ can be considered uniform, and thus sum together
all the contributions~(\ref{eq:intens-hom}) of the different segments.
The dynamical structure factor of the inhomogeneous system
would then be approximated by:
\begin{eqnarray}
  \label{eq:approx-tosi}
  S(q,\omega)=\int \frac{dx}{2R} S_{\mathrm{hom}}(q,\omega,n_0(x))\; .
\end{eqnarray}
$S_{\mathrm{hom}}(q,\omega,n)$ is given by
Eq.~(\ref{eq:intens-hom}), where now the Tomonaga-Luttinger
parameters $u=u(n)$ and $K=K(n)$ depend on density.

With the help of (\ref{eq:intens-hom}), we obtain:
\begin{eqnarray}
  S(q,\omega)&&= \frac {|q|} {4R} \int_{-R}^R dx K(n_0(x))\\
 && [\delta(\omega - u(n_0(x)) |q|) +  \delta(\omega + u(n_0(x))
    |q|) ]\nonumber
\end{eqnarray}
Introducing $x^*(\omega/|q|)$, such that $\omega = u(n(x^*))|q|$ we
can rewrite:
\begin{eqnarray}
  \label{eq:s-lda}
  S(q,\omega)=\frac{K(n_0(x^*))}{2R \left|\frac{du}{dn}\right|_{n=n_0(x^*)}
    \left| \frac {d n_0}{dx}\right|_{x=x*}}
\end{eqnarray}
Since the compressibility is a positive quantity, the chemical
potential is an increasing function of the density. Moreover
for a trapping potential that is an increasing function
of position, from Eq.~(\ref{eq:thomas-fermi_1})) the density is seen to 
decrease with position. Thus, when the velocity is an increasing 
function of density, the solution $x^*$ turns out to be unique

The quantity $\frac {d n_0}{dx}$ can be obtained by differentiating the
relation~(\ref{eq:thomas-fermi_1}) with respect to $x$ , i.e.:
\begin{eqnarray}
  \left(\frac{d^2 e}{dn^2}\right)_{n=n_0(x)} \frac{d n_0}{dx} +
  \frac{dV}{dx}=0\; .
\end{eqnarray}
We can therefore write:
\begin{eqnarray}
\label{eq:Sqw}
   S(q,\omega)=\frac{K(n_0(x^*))\left|\frac{d^2e}{dn^2}\right|_{n=n_0(x^*)}}{2R\left|\frac{du}{dn}\right|_{n=n_0(x^*)}
     \left| \frac {dV}{dx}\right|_{x=x*}}.
\end{eqnarray}

We now use the relation $u^2(n)=\frac n m \frac{d^2 e}{dn^2}$ obtained
from (\ref{eq:uK}) and rewrite (\ref{eq:Sqw}) as:
\begin{eqnarray}
  \label{eq:s-lda-master}
  S(q,\omega)=\frac{\pi \hbar}{R \left|\frac {dV}{dx}\right|_{x=x^*}}
  \frac{n_0(x^*)}{\left|1+n_0(x^*) \frac
    {e'''(n_0(x^*))}{e''(n_0(x^*))}\right|},
\end{eqnarray}
with the notations $e'(x)=de/dn$, $e''(n)=d^2e/dn^2$, and
$e'''(n)=d^3e/dn^3$.

Formula (\ref{eq:s-lda-master}) represents the main result of this
paper. It gives an analytical expression for the light scattering
cross-section of an inhomogeneous TLL once the ground state energy as a
function of the density is known, e.g. by an exact analytical
(Bethe-Ansatz) or via numerical simulations (QMC). Remarkably,
Eq.~(\ref{eq:s-lda-master}) predicts that $S(q,\omega)$ is only a function of
$\omega/|q|$. In fact, this is the specific signature of
Tomonaga-Luttinger Liquid behavior in shallow trapped 1D Bose systems, as it
can be measured by Bragg spectroscopy.

In order to illustrate the relevant features and make the connection
with Eq. (\ref{eq:final-boso}) obtained {\it via}
the hydrodynamic approach of Sec.~\ref{sec:hydro}, we now treat the
case of harmonic trapping. In this case
$dV/dx=m\Omega_0^2 x$, and using Eq.~(\ref{eq:thomas-fermi_1}) we have
$m\Omega_0^2 |x^*|=\sqrt{2 m\Omega_0^2(e'(n_0(0))-e'(\rho^*))}$,
where we have set $\rho^\star=n_0(x^\star)$ and $u(\rho^*)=\omega/|q|$.
Eq. (\ref{eq:s-lda-master}) thus simplifies into:
\begin{eqnarray}
  \label{eq:s-lda-harmonic}
  S(q,\omega)=\frac{\pi \hbar \rho^*}{R \sqrt{2 m \Omega_0^2
      (e'(n_0(0))-e'(\rho^*))}\left|1 + \rho^*
    \frac{e'''(\rho^*)}{e''(\rho_*)}\right|}.
      \end{eqnarray}

We now check the consistency of the result (\ref{eq:s-lda-master})
with (\ref{eq:final-boso}), by explicitly calculating (\ref{eq:s-lda-harmonic})
for the model $e(n)\propto n^{\gamma
  +2}$. Eq.~(\ref{eq:s-lda-harmonic}) then reads:
\begin{eqnarray}\label{eq:power-case}
  S(q,\omega)=\frac{\pi \hbar}{(\gamma+1)m\Omega_0^2 R^2} \left(\frac{m
    \Omega_0^2 R^2}{2g(\gamma+2)}\right)^{\frac 1 {\gamma+1}}
  \frac{\left(\frac{\omega}{u_0 q}\right)^{\frac 2
      {\gamma+1}}}{\sqrt{1-\left(\frac{\omega}{u_0 q}\right)^2}},
\end{eqnarray}
where we have defined $u_0=u(n_0(0))$ as the velocity of excitations in
a uniform system having  a density equal to that at the trap center.
We first notice that the dynamical structure factor in
(\ref{eq:power-case}) makes explicit the characteristic already
embodied in the structure of Eq. (\ref{eq:s-lda-harmonic}), namely
that $S(q,\omega)$ depends on wavevector and frequency
solely through their ratio
$\omega/q$. Second, the formula (\ref{eq:power-case})
with $\gamma=1$ agrees with the result of
Ref.~\cite{vignolo2001}, in the limiting $\omega \gg q^2/2$
case. Finally, in
App. \ref{sec:sums} we show by inspection
that the LDA approximation (\ref{eq:power-case}) is fully recovered
from expression (\ref{eq:final-boso}).

\begin{figure}[t]
\centering \includegraphics[scale=0.2]{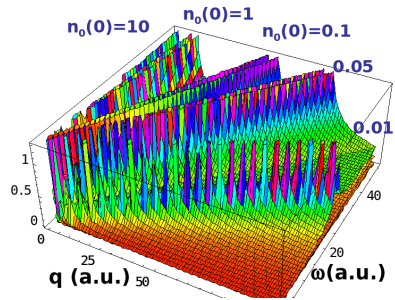}
\caption{(Color online).
TLL model with $e(nr_0)\propto n^\gamma$ and $\gamma=2$
in a harmonic trap.
$S(q,\omega)$ in arbitrary units in the ($\omega, q$) plane
  and different densities at the trap center.}
\label{fig:bragg3d-plot}
\end{figure}

Fig.~\ref{fig:bragg3d-plot} displays the 3D plot of $S(q,\omega)$
resulting from the use of
(\ref{eq:power-case}) in the $(\omega,q)$ plane, while varying the
densities at the trap center. $S(q,\omega)$ is a set of discrete peaks
whose position varies linearly with $\omega/q$ and such linear
behavior is independent on the interaction strength.

Before proceeding to apply Eq. (\ref{eq:s-lda-harmonic}) to a dipolar
1D Bose gas, we step on commenting
the found correspondence between
hydrodynamic and  DFT-LDA approaches on a more general footing.
It is well known for normal Fermi
systems~\cite{Vignale97} with extension to Bose
superfluids~\cite{Chiofalo98}, that the treatment of dynamical
processes in interacting inhomogeneous  systems do
require the development the Current-Density Functional Theory, where
invariance conditions render the energy to be a functional of the
current besides density. It was demonstrated that the analogue of LDA
leads in this case to Navier-Stokes equations
(Landau-Khalatnikov two-fluid equations for superfluids),
where viscosities, densities and currents (normal and superfluid)
have a microscopic expression in terms of Kubo relations
and low-frequency response functions as they can be
calculated in the homogeneous system at the local densities and
currents. Such a general view is reflected by the present result.
In the Tomonaga-Luttinger-Liquid free harmonic Hamiltonian,
where the interactions are effectively embodied in $u$ and $K$, the
Navier-Stokes equations become indeed the simple hydrodynamic relations of
Sec.~\ref{sec:hydro}. On the other hand, in the DFT and LDA approach of
Sec.~\ref{sec:approx} the treatment explicitly uses the
two mappings: from interacting to non-interacting system (DFT) and
from inhomogeneous to homogeneous (LDA).

\section{1D Bose gases coupled {\it via} dipolar interactions}
~\label{sec:TLLdip}

In this Section we specialize to the case of a 1D dipolar gas in a
harmonic trapping potential. We first recall the main results known
for the homogeneous system, and then apply Eq. (\ref{eq:s-lda-harmonic})
to determine the scattered light intensity. The system is characterized by the
strength of the interactions $C_{dd}$, resulting from either magnetic
$C_{dd}=\mu_0\mu_d^2$ or electric $C_{dd}=d^2/\epsilon_0$
dipoles, where $\mu_d$ and $d$ are the magnetic and electric dipole
  moments and $\mu_0$ and $\epsilon_0$ are the vacuum
  permittivities. An effective Bohr radius can be defined from
$C_{dd}$ as $r_0\equiv M C_{dd}/(2 \pi \hbar^2)$ and  the Hamiltonian
in effective Rydberg units $Ry^*={\hbar^2}/({2 M r_0^2})$ is
\begin{equation}
H=(n r_0)^2 \left[ -\sum_{i} \frac{\partial^2}{\partial x_i^2}+ (n
  r_0) \sum_{i<j}\frac{1}{|x_i-x_j|^3} \right] \; ,
\label{eq:HdipoleRy}
\end{equation}
where lenghts are expressed in $1/n$ units.
The physics of the model is entirely specified by the dimensionless
coupling parameter $n r_0$, so that in the high-density limit the
system becomes strongly correlated and a quasi-ordered state occurs,
where the potential energy dominates.

The ground-state energy $e(n)$
of this model was determined by means of Reptation QMC method  in
Ref.~\cite{ours}. In the low
$nr_0\to 0$ limit
it reproduces the Tonks-Girardeau (TG) state
energy per particle of a
free spinless Fermi gas, whose energy per particle is
$e_{TG}(n)=\pi^2(nr_0)^2/3$ $Ry^*$.
In the large
$nr_0\to \infty$ limit of high-density dipoles, it reproduces
the Dipolar Density Wave (DDW) state  where
$e_{DDW}(n)=\zeta(3) (n r_0)^3$ $Ry^*$ and
$\zeta(3)=1.20205$.
The QMC thermodynamic energy per particle in Rydberg units can be
represented as an analytical function of $n r_0$:
\begin{eqnarray}
&&  e_p(nr_0)=\frac{\zeta(3) (nr_0)^4+a (nr_0)^e+ b (nr_0)^f + c
   (nr_0)^{(2+g)}}{1+nr_0} \\
&& +\frac{\pi^2}{3} \frac{(n r_0)^2}{1+d
   (nr_0)^g}\ \nonumber \; .
\label{eq:fit}
\end{eqnarray}
The fitting coefficients, yielding a reduced $\chi^2 \simeq 5$, are:
$a =3.1(1)$, $b = 3.2(2)$, $c = 4.3(4)$, $d=  1.7(1)$, $e=  3.503(4)$,
$f=  3.05(5)$, and $g=  0.34(4)$.

The Bragg intensity is thus easily
obtained by Eq.~(\ref{eq:s-lda-harmonic}) once the value of
$\rho^\star$ is determined.
\begin{figure}[h]
\centering \includegraphics[scale=0.7]{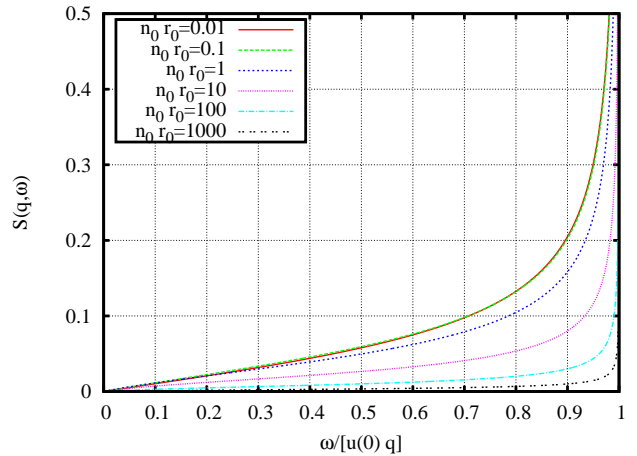}
\caption{(Color online).
1D dipolar Bose gas confined in an harmonic trap,
with $e(n)$ as determined by
  QMC simulations.
$S(q,\omega)$ (arbitrary units) {\it vs.} $\omega/(u_0 q)$
for different densities $n_0$ at the trap center.
The values of $n_0 r_0$ running from 0.01 to 1000 are indicated in the legend.}
\label{fig:scaled-plot}
\end{figure}
In Fig.\ref{fig:scaled-plot} we report the scaling behavior of
$S(q,\omega)$ {\it vs.} $\omega/(u_0 q)$ for different densities at the
trap center $n_0$. Larger $n_0 r_0$ indicate stronger coupling
interactions, crossing over from TG to DDW states.
The linear behavior in the low $\omega/q$ regime is
striking, the slope continuously increasing with decreasing $u(0)$ and
thus $n_0$.  In the TG limit, the tail of
$S(q,\omega)$ is insensitive to changes of the density at the
center of the trap, and in fact the curves with $n_0r_0=0.01$ and
$0.1$ do coincide. The comparison with the TG gas ($\gamma=1$ and
$n_0r_0=0.01$) and the DDW case ($\gamma=2$ and
$n_0r_0=10^3$) is better seen in Fig.\ref{fig:crossover},  where
$S(q,\omega) \sqrt{e'(n_0)}/n_0$ is plotted as a function of
$\omega/(u_0 q)$. One can notice that a crossover takes place in the
intermediate densities regime. Viewed in the log-log scale, the plot
evidences how a measure of
the $S(q,\omega)$ tail towards small $\omega/q$, would
provide a way to determine the interaction regime. A peculiarity of
the TLL behavior is the power-law trend when $\omega/(u_0q=1)$
is approached. A detailed study of the power-law non-analyticity for a
trapped Bose gas can be found in
~\cite{golovach_2009}.

\begin{figure}[h]
\centering \includegraphics[scale=0.7]{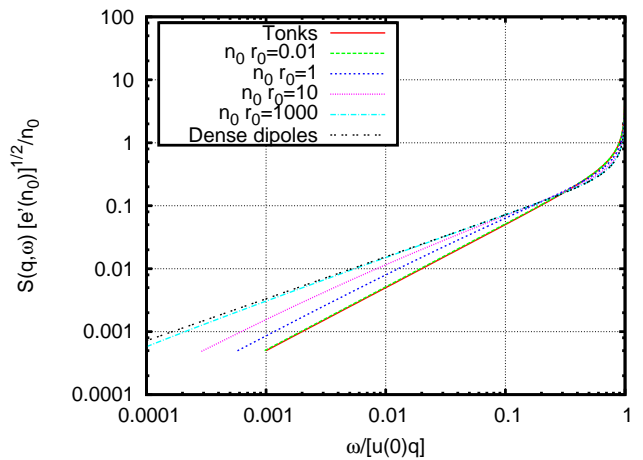}
\caption{The same as Fig.~\ref{fig:scaled-plot}, but in a log-log
  scale. The
  comparison with the Tonks-Girardeau limit
gas ($\gamma=1$) and the dense dipole
  limit corresponding to a Dipolar Density Wave ($\gamma=2$) is shown
  in evident manner.}
\label{fig:crossover}
\end{figure}

\section{Conclusions}~\label{sec:conc}
We have derived the dynamical structure factor for an
inhomogeneous Tomonaga-Luttinger liquid as it can occur in a confined
strongly interacting one-dimensional gas. In view of current
experimental progress in the field, we have provided an
easy-to-use and simple analytical expression for the light-scattering cross
section, Eq. (\ref{eq:s-lda-master}), valid within a Local Density
Approximation.

The analytical expression (\ref{eq:s-lda-master})
predicts that $S(q,\omega)$ is only a function of
$\omega/|q|$ and is the central result of this work.
In fact, this is the specific signature of
Tomonaga-Luttinger Liquid behavior in shallow trapped 1D Bose systems, along
with a power-law behavior when $\omega/(u_0 q)$ is approached, as it
can be measured by Bragg spectroscopy.

Expression (\ref{eq:s-lda-master}) is validated by the
independent derivation (\ref{eq:final-boso}) by means of a
hydrodynamic approach, which is reported in detail in
App.~\ref{sec:sums}. The connection between the two approaches is
a second result of this work, and is a consequence of the more
general Current-Density
Functional Theory~\cite{Vignale97,Chiofalo98} applied to the conditions
of the present work.

We thus remark that expression (\ref{eq:s-lda-master}) can be in principle
applied to the many 1D systems cited in the introductory
material, once the trapping potential is known together with
the ground state energy as a
function of the density, {\it e.g.} by means of perturbative, exact, or
computational methods applied to the homogeneous system. Extension of
the present method to include additional local perturbations coupling
to the density, could be used to investigate the propagation of local
density fluctuations. 

Finally, we have applied our findings to the case of
one-dimensional quantum bosonic gases with dipolar interactions,
using the harmonic profile typical of
experiments in this field, accompanied by
our previous QMC
data for the energy per particle. We find an universal
scaling behavior peculiar of the Tomonaga-Luttinger
liquid~\cite{ours}, a signature that
can be eventually probed by Bragg spectroscopy in ongoing
experimental realizations of such systems~\cite{Chotia_2011}.

\appendix
\section{Justification of the LDA formula}
\label{sec:sums}

In order to justify the approximate formulas, it is more convenient to
work with the integrated intensity:
\begin{eqnarray}
  \label{eq:integrated}
  I(q,\Omega)=\int_0^\Omega S(q,\omega)d\omega\; ,
\end{eqnarray}
since the delta functions in the sum~(\ref{eq:final-boso})
contribute as step functions in $I(q,\Omega)$ giving more regular
expressions.

Using the approximation~(\ref{eq:power-case}), we expect:
\begin{eqnarray}
\nonumber
&& I(q,\Omega)=\frac{\pi \hbar u_0 n_0 |q|}{2m \Omega_0^2 R^2}
 \frac{2\alpha+1}{\alpha+1} \left(\frac{\Omega}{u_0
     q}\right)^{2(\alpha+1)}\\
&& \times  {}_2F_1\left[\alpha+1,\frac 1
   2;\alpha+2;\left(\frac{\Omega}{u_0 q}\right)^2\right]\; ,
 \label{eq:integ-approx}
\end{eqnarray}
where ${}_2F_1$ is the Gauss hypergeometric function.
Using the expression~(\ref{eq:final-boso}), we obtain instead the
exact expression:

\begin{eqnarray}
  \label{eq:integ-exact}
  I(q,\omega_N)&\propto& \sum_{n<N} \frac{2(n+\alpha+1)
    \Gamma(n+2\alpha+2)}{n!}\\
&& \times \frac{J^2_{n+\alpha+1}(qR)}{\sqrt{(n+1)(n+2\alpha+1)}}\; ,\nonumber
\end{eqnarray}
where $N$ is such that $\omega_N=\Omega$. In order to check the
consistency between (\ref{eq:integ-approx}) and
(\ref{eq:integ-exact}) we can work on the sums in
(\ref{eq:integ-exact}). We expect that the sum is dominated by the
terms having $n\gg 1$. Using Eq. (9.3.2) in
Ref.~\cite{abramowitz_math_functions}, we expect that for
$qR<n$, $J^2_{n+\alpha+1}(qR)$ is an exponentially small quantity with
$n$. For $qR>n$ however, Eq. (9.3.3) in
Ref.~\cite{abramowitz_math_functions} suggests that:
\begin{eqnarray}
  J^2_{n+\alpha+1}(qR)&\simeq& \frac 2 {\pi (n+\alpha+1) \tan \beta}\\
&&  \times \cos^2((n+\alpha+1)(\tan \beta -\beta)-\pi/4)\; ,\nonumber
\end{eqnarray}
where $\cos \beta = qR/(n+\alpha+1)$. Elementary trigonometry gives
the approximation:
\begin{eqnarray}
\label{eq:approx-bes-2}
  J^2_{n+\alpha+1}(qR)&\simeq& \frac{(n+\alpha+1)}{\pi
    \sqrt{(qR)^2-(n+\alpha+1)^2}}\\
&&\times \left\{1+\sin \left[2
  \sqrt{(qR)^2-(n+\alpha+1)^2}\right.\right. \nonumber\\ &&
\left.\left. -(n+\alpha+1)
\arccos
       \left(\frac{n+\alpha+1}{qR}\right) \right]\right\}\nonumber
\end{eqnarray}
Dropping the term oscillating with $n$ in (\ref{eq:approx-bes-2}),
we use as approximation:
\begin{eqnarray}
  \label{eq:crude}
  J^2_{n+\alpha+1}(qR) \simeq \frac{\theta(qR - n -\alpha -1)}{\pi
    \sqrt{(qR)^2-(n+\alpha+1)^2}}
\end{eqnarray}
We can also approximate:
\begin{eqnarray}
  \frac{\Gamma(n+2\alpha+2)}{\Gamma(n+1)} &\simeq& n^{2\alpha+1}
  \\ \frac{n+\alpha+1}{\sqrt{(n+1){n+2\alpha+1}}} &\simeq& 1
\end{eqnarray}
So that the sum in Eq.~(\ref{eq:integ-exact}) can be approximated by:
\begin{eqnarray}
\label{eq:approx-sum}
  \frac 2 \pi \sum_{n<\min(N,qR)}
  \frac{n^{2\alpha+1}}{\sqrt{(qR)^2-(n+\alpha+1)^2}}
\end{eqnarray}
Finally, by approximating the sum~(\ref{eq:approx-sum})
by an integral, we find:
\begin{eqnarray}
   &&\sum_{n<\min(N,qR)}
   \frac{n^{2\alpha+1}}{\sqrt{(qR)^2-(n+\alpha+1)^2}} \simeq \nonumber
   \\
   && \simeq \int_0^{\min(N,qR)} \frac{du u^{2\alpha+1}}{\sqrt{(qR)^2-u^2}}
\end{eqnarray}
Using Eq. (6.6.1) in \cite{abramowitz_math_functions}, we have (for
$qR<N$):
\begin{eqnarray}
\label{crude-beta}
  \int_0^{\min(N,qR)} \frac{du u^{2\alpha+1}}{\sqrt{(qR)^2-u^2}} \nonumber
   \\
   = \frac 1 2 (qR)^{2\alpha+1} B_{(\Omega/qR\Omega_0)^2}(\alpha+1,1/2),
\end{eqnarray}
where $B_x(a,b)$ are the incomplete Beta functions.
With Eq. (6.6.8) of~\cite{abramowitz_math_functions} we can
check that Eq.~(\ref{eq:integ-approx}) agrees with the obtained
approximate expression~(\ref{crude-beta}).

\begin{acknowledgments} MLC would like to thank Debbie Jin for
  illuminating discussions on the experiments ~\cite{Chotia_2011} and
  their potential follow-up.
\end{acknowledgments}


\begin{thebibliography}{64}
\expandafter\ifx\csname natexlab\endcsname\relax\def\natexlab#1{#1}\fi
\expandafter\ifx\csname bibnamefont\endcsname\relax
  \def\bibnamefont#1{#1}\fi
\expandafter\ifx\csname bibfnamefont\endcsname\relax
  \def\bibfnamefont#1{#1}\fi
\expandafter\ifx\csname citenamefont\endcsname\relax
  \def\citenamefont#1{#1}\fi
\expandafter\ifx\csname url\endcsname\relax
  \def\url#1{\texttt{#1}}\fi
\expandafter\ifx\csname urlprefix\endcsname\relax\def\urlprefix{URL }\fi
\providecommand{\bibinfo}[2]{#2}
\providecommand{\eprint}[2][]{\url{#2}}

\bibitem[{\citenamefont{Varma et~al.}(2002)\citenamefont{Varma, Nussinov, and
  {van Saarloos}}}]{varma_nonfermiliquid_review}
\bibinfo{author}{\bibfnamefont{C.~M.} \bibnamefont{Varma}},
  \bibinfo{author}{\bibfnamefont{Z.}~\bibnamefont{Nussinov}}, \bibnamefont{and}
  \bibinfo{author}{\bibfnamefont{W.}~\bibnamefont{{van Saarloos}}},
  \bibinfo{journal}{Phys. Rep.} \textbf{\bibinfo{volume}{361}},
  \bibinfo{pages}{267} (\bibinfo{year}{2002}).

\bibitem[{\citenamefont{Schulz}(1995)}]{schulz_houches_revue}
\bibinfo{author}{\bibfnamefont{H.~J.} \bibnamefont{Schulz}}, in
  \emph{\bibinfo{booktitle}{Mesoscopic Quantum Physics, Les Houches LXI}},
  edited by \bibinfo{editor}{\bibfnamefont{E.}~\bibnamefont{Akkermans}},
  \bibinfo{editor}{\bibfnamefont{G.}~\bibnamefont{Montambaux}},
  \bibinfo{editor}{\bibfnamefont{J.~L.} \bibnamefont{Pichard}},
  \bibnamefont{and}
  \bibinfo{editor}{\bibfnamefont{J.}~\bibnamefont{Zinn-Justin}}
  (\bibinfo{publisher}{Elsevier}, \bibinfo{address}{Amsterdam},
  \bibinfo{year}{1995}), p. \bibinfo{pages}{533}.

\bibitem[{\citenamefont{Bourbonnais and
  J{\'e}rome}(1999)}]{bourbonnais_jerome_review}
\bibinfo{author}{\bibfnamefont{C.}~\bibnamefont{Bourbonnais}} \bibnamefont{and}
  \bibinfo{author}{\bibfnamefont{D.}~\bibnamefont{J{\'e}rome}}, in
  \emph{\bibinfo{booktitle}{Advances in Synthetic Metals, Twenty years of
  Progress in Science and Technology}}, edited by
  \bibinfo{editor}{\bibfnamefont{P.}~\bibnamefont{Bernier}},
  \bibinfo{editor}{\bibfnamefont{S.}~\bibnamefont{Lefrant}}, \bibnamefont{and}
  \bibinfo{editor}{\bibfnamefont{G.}~\bibnamefont{Bidan}}
  (\bibinfo{publisher}{Elsevier}, \bibinfo{address}{New York},
  \bibinfo{year}{1999}), p. \bibinfo{pages}{206}, \bibinfo{note}{preprint
  cond-mat/9903101}.

\bibitem[{\citenamefont{Bourbonnais and Jerome}(2008)}]{bourbonnais2008}
\bibinfo{author}{\bibfnamefont{C.}~\bibnamefont{Bourbonnais}} \bibnamefont{and}
  \bibinfo{author}{\bibfnamefont{D.}~\bibnamefont{Jerome}}, in
  \emph{\bibinfo{booktitle}{Physics of Organic Superconductors and
  Conductors}}, edited by \bibinfo{editor}{\bibfnamefont{A.~G.}
  \bibnamefont{Lebed}} (\bibinfo{publisher}{Springer}, \bibinfo{year}{2008}),
  vol. \bibinfo{volume}{110} of \emph{\bibinfo{series}{Springer Series in
  Materials Science}}, p. \bibinfo{pages}{357},
  \bibinfo{note}{arXiv:0904.0617}.

\bibitem[{\citenamefont{Mizokawa et~al.}(2002)\citenamefont{Mizokawa, Nakada,
  Kim, Shen, Yoshida, Fujimori, Horii, Yamada, Ikuta, and
  Mizutani}}]{mizokawa2002}
\bibinfo{author}{\bibfnamefont{T.}~\bibnamefont{Mizokawa}},
  \bibinfo{author}{\bibfnamefont{K.}~\bibnamefont{Nakada}},
  \bibinfo{author}{\bibfnamefont{C.}~\bibnamefont{Kim}},
  \bibinfo{author}{\bibfnamefont{Z.-X.} \bibnamefont{Shen}},
  \bibinfo{author}{\bibfnamefont{T.}~\bibnamefont{Yoshida}},
  \bibinfo{author}{\bibfnamefont{A.}~\bibnamefont{Fujimori}},
  \bibinfo{author}{\bibfnamefont{S.}~\bibnamefont{Horii}},
  \bibinfo{author}{\bibfnamefont{Y.}~\bibnamefont{Yamada}},
  \bibinfo{author}{\bibfnamefont{H.}~\bibnamefont{Ikuta}}, \bibnamefont{and}
  \bibinfo{author}{\bibfnamefont{U.}~\bibnamefont{Mizutani}},
  \bibinfo{journal}{Phys. Rev. B} \textbf{\bibinfo{volume}{65}},
  \bibinfo{pages}{193101} (\bibinfo{year}{2002}).

\bibitem[{\citenamefont{Wang et~al.}(2006)\citenamefont{Wang, Mo, Allen, Kim,
  He, Jin, Mandrus, Sekiyama, Tsunekawa, and Suga}}]{wang06_arpes_limo6o17}
\bibinfo{author}{\bibfnamefont{F.}~\bibnamefont{Wang}},
  \bibinfo{author}{\bibfnamefont{S.-K.} \bibnamefont{Mo}},
  \bibinfo{author}{\bibfnamefont{J.~W.} \bibnamefont{Allen}},
  \bibinfo{author}{\bibfnamefont{H.-D.} \bibnamefont{Kim}},
  \bibinfo{author}{\bibfnamefont{J.}~\bibnamefont{He}},
  \bibinfo{author}{\bibfnamefont{R.}~\bibnamefont{Jin}},
  \bibinfo{author}{\bibfnamefont{D.}~\bibnamefont{Mandrus}},
  \bibinfo{author}{\bibfnamefont{A.}~\bibnamefont{Sekiyama}},
  \bibinfo{author}{\bibfnamefont{M.}~\bibnamefont{Tsunekawa}},
  \bibnamefont{and} \bibinfo{author}{\bibfnamefont{S.}~\bibnamefont{Suga}},
  \bibinfo{journal}{Phys. Rev. B} \textbf{\bibinfo{volume}{74}},
  \bibinfo{pages}{113107} (\bibinfo{year}{2006}).

\bibitem[{\citenamefont{{Hammar} et~al.}(1999)\citenamefont{{Hammar}, {Stone},
  {Reich}, {Broholm}, {Gibson}, {Turnbull}, {Landee}, and
  {Oshikawa}}}]{hammar1999}
\bibinfo{author}{\bibfnamefont{P.~R.} \bibnamefont{{Hammar}}},
  \bibinfo{author}{\bibfnamefont{M.~B.} \bibnamefont{{Stone}}},
  \bibinfo{author}{\bibfnamefont{D.~H.} \bibnamefont{{Reich}}},
  \bibinfo{author}{\bibfnamefont{C.}~\bibnamefont{{Broholm}}},
  \bibinfo{author}{\bibfnamefont{P.~J.} \bibnamefont{{Gibson}}},
  \bibinfo{author}{\bibfnamefont{M.~M.} \bibnamefont{{Turnbull}}},
  \bibinfo{author}{\bibfnamefont{C.~P.} \bibnamefont{{Landee}}},
  \bibnamefont{and}
  \bibinfo{author}{\bibfnamefont{M.}~\bibnamefont{{Oshikawa}}},
  \bibinfo{journal}{Phys. Rev. B} \textbf{\bibinfo{volume}{59}},
  \bibinfo{pages}{1008} (\bibinfo{year}{1999}),
  \eprint{arXiv:cond-mat/9809068}.

\bibitem[{\citenamefont{{Lake} et~al.}(2005)\citenamefont{{Lake}, {Tennant},
  {Frost}, and {Nagler}}}]{lake2005}
\bibinfo{author}{\bibfnamefont{B.}~\bibnamefont{{Lake}}},
  \bibinfo{author}{\bibfnamefont{D.~A.} \bibnamefont{{Tennant}}},
  \bibinfo{author}{\bibfnamefont{C.~D.} \bibnamefont{{Frost}}},
  \bibnamefont{and} \bibinfo{author}{\bibfnamefont{S.~E.}
  \bibnamefont{{Nagler}}}, \bibinfo{journal}{Nat. Mater.}
  \textbf{\bibinfo{volume}{4}}, \bibinfo{pages}{329} (\bibinfo{year}{2005}),
  \eprint{arXiv:cond-mat/0501069}.

\bibitem[{\citenamefont{Dagotto}(1999)}]{dagotto_supra_ladder_review}
\bibinfo{author}{\bibfnamefont{E.}~\bibnamefont{Dagotto}},
  \bibinfo{journal}{Rep. Prog. Phys.} \textbf{\bibinfo{volume}{62}},
  \bibinfo{pages}{1525} (\bibinfo{year}{1999}).

\bibitem[{\citenamefont{{Klanjsek} et~al.}(2008)\citenamefont{{Klanjsek},
  {Mayaffre}, {Berthier}, {Horvatic}, {Chiari}, {Piovesana}, {Bouillot},
  {Kollath}, {Orignac}, {Citro} et~al.}}]{klanjsek08_bpcp}
\bibinfo{author}{\bibfnamefont{M.}~\bibnamefont{{Klanjsek}}},
  \bibinfo{author}{\bibfnamefont{H.}~\bibnamefont{{Mayaffre}}},
  \bibinfo{author}{\bibfnamefont{C.}~\bibnamefont{{Berthier}}},
  \bibinfo{author}{\bibfnamefont{M.}~\bibnamefont{{Horvatic}}},
  \bibinfo{author}{\bibfnamefont{B.}~\bibnamefont{{Chiari}}},
  \bibinfo{author}{\bibfnamefont{O.}~\bibnamefont{{Piovesana}}},
  \bibinfo{author}{\bibfnamefont{P.}~\bibnamefont{{Bouillot}}},
  \bibinfo{author}{\bibfnamefont{C.}~\bibnamefont{{Kollath}}},
  \bibinfo{author}{\bibfnamefont{E.}~\bibnamefont{{Orignac}}},
  \bibinfo{author}{\bibfnamefont{R.}~\bibnamefont{{Citro}}},
  \bibnamefont{et~al.}, \bibinfo{journal}{Phys. Rev. Lett.}
  \textbf{\bibinfo{volume}{101}}, \bibinfo{pages}{137207}
  (\bibinfo{year}{2008}), \bibinfo{note}{arXiv:0804.2639}.

\bibitem[{\citenamefont{Auslaender et~al.}(2002)\citenamefont{Auslaender,
  Yacoby, {de Picciotto}, Baldwin, Pfeiffer, and
  West}}]{auslaender_quantumwire_tunneling}
\bibinfo{author}{\bibfnamefont{O.~M.} \bibnamefont{Auslaender}},
  \bibinfo{author}{\bibfnamefont{A.}~\bibnamefont{Yacoby}},
  \bibinfo{author}{\bibfnamefont{R.}~\bibnamefont{{de Picciotto}}},
  \bibinfo{author}{\bibfnamefont{K.~W.} \bibnamefont{Baldwin}},
  \bibinfo{author}{\bibfnamefont{L.~N.} \bibnamefont{Pfeiffer}},
  \bibnamefont{and} \bibinfo{author}{\bibfnamefont{K.~W.} \bibnamefont{West}},
  \bibinfo{journal}{Science} \textbf{\bibinfo{volume}{295}},
  \bibinfo{pages}{825} (\bibinfo{year}{2002}).

\bibitem[{\citenamefont{{Hilke} et~al.}(2001)\citenamefont{{Hilke}, {Tsui},
  {Grayson}, {Pfeiffer}, and {West}}}]{hilke2001}
\bibinfo{author}{\bibfnamefont{M.}~\bibnamefont{{Hilke}}},
  \bibinfo{author}{\bibfnamefont{D.~C.} \bibnamefont{{Tsui}}},
  \bibinfo{author}{\bibfnamefont{M.}~\bibnamefont{{Grayson}}},
  \bibinfo{author}{\bibfnamefont{L.~N.} \bibnamefont{{Pfeiffer}}},
  \bibnamefont{and} \bibinfo{author}{\bibfnamefont{K.~W.}
  \bibnamefont{{West}}}, \bibinfo{journal}{Phys. Rev. Lett.}
  \textbf{\bibinfo{volume}{87}}, \bibinfo{pages}{186806}
  (\bibinfo{year}{2001}), \eprint{arXiv:cond-mat/0104553}.

\bibitem[{\citenamefont{Bockrath et~al.}(1999)\citenamefont{Bockrath, Cobden,
  Lu, Rinzler, Smalley, Balents, and Mceuen}}]{bockrath_luttinger_nanotubes}
\bibinfo{author}{\bibfnamefont{M.}~\bibnamefont{Bockrath}},
  \bibinfo{author}{\bibfnamefont{D.~H.} \bibnamefont{Cobden}},
  \bibinfo{author}{\bibfnamefont{J.}~\bibnamefont{Lu}},
  \bibinfo{author}{\bibfnamefont{A.~G.} \bibnamefont{Rinzler}},
  \bibinfo{author}{\bibfnamefont{R.~E.} \bibnamefont{Smalley}},
  \bibinfo{author}{\bibfnamefont{L.}~\bibnamefont{Balents}}, \bibnamefont{and}
  \bibinfo{author}{\bibfnamefont{P.~L.} \bibnamefont{Mceuen}},
  \bibinfo{journal}{Nature (London)} \textbf{\bibinfo{volume}{397}},
  \bibinfo{pages}{598} (\bibinfo{year}{1999}).

\bibitem[{\citenamefont{Ishii et~al.}(2003)\citenamefont{Ishii, Kataura,
  Shiozawa, Yoshioka, Otsubo, Takayama, Miyahara, Suzuki, Achiba, Nakatake
  et~al.}}]{ishii03_nanotube_pes}
\bibinfo{author}{\bibfnamefont{H.}~\bibnamefont{Ishii}},
  \bibinfo{author}{\bibfnamefont{H.}~\bibnamefont{Kataura}},
  \bibinfo{author}{\bibfnamefont{H.}~\bibnamefont{Shiozawa}},
  \bibinfo{author}{\bibfnamefont{H.}~\bibnamefont{Yoshioka}},
  \bibinfo{author}{\bibfnamefont{H.}~\bibnamefont{Otsubo}},
  \bibinfo{author}{\bibfnamefont{Y.}~\bibnamefont{Takayama}},
  \bibinfo{author}{\bibfnamefont{T.}~\bibnamefont{Miyahara}},
  \bibinfo{author}{\bibfnamefont{S.}~\bibnamefont{Suzuki}},
  \bibinfo{author}{\bibfnamefont{Y.}~\bibnamefont{Achiba}},
  \bibinfo{author}{\bibfnamefont{M.}~\bibnamefont{Nakatake}},
  \bibnamefont{et~al.}, \bibinfo{journal}{Nature (London)}
  \textbf{\bibinfo{volume}{426}}, \bibinfo{pages}{540} (\bibinfo{year}{2003}).

\bibitem[{\citenamefont{{Gao} et~al.}(2004)\citenamefont{{Gao}, {Komnik},
  {Egger}, {Glattli}, and {Bachtold}}}]{gao2004}
\bibinfo{author}{\bibfnamefont{B.}~\bibnamefont{{Gao}}},
  \bibinfo{author}{\bibfnamefont{A.}~\bibnamefont{{Komnik}}},
  \bibinfo{author}{\bibfnamefont{R.}~\bibnamefont{{Egger}}},
  \bibinfo{author}{\bibfnamefont{D.~C.} \bibnamefont{{Glattli}}},
  \bibnamefont{and}
  \bibinfo{author}{\bibfnamefont{A.}~\bibnamefont{{Bachtold}}},
  \bibinfo{journal}{Phys. Rev. Lett.} \textbf{\bibinfo{volume}{92}},
  \bibinfo{pages}{216804} (\bibinfo{year}{2004}),
  \eprint{arXiv:cond-mat/0311645}.

\bibitem[{\citenamefont{{Lee} et~al.}(2004)\citenamefont{{Lee}, {Eggert},
  {Kim}, {Kahng}, {Shinohara}, and {Kuk}}}]{lee2004}
\bibinfo{author}{\bibfnamefont{J.}~\bibnamefont{{Lee}}},
  \bibinfo{author}{\bibfnamefont{S.}~\bibnamefont{{Eggert}}},
  \bibinfo{author}{\bibfnamefont{H.}~\bibnamefont{{Kim}}},
  \bibinfo{author}{\bibfnamefont{S.-J.} \bibnamefont{{Kahng}}},
  \bibinfo{author}{\bibfnamefont{H.}~\bibnamefont{{Shinohara}}},
  \bibnamefont{and} \bibinfo{author}{\bibfnamefont{Y.}~\bibnamefont{{Kuk}}},
  \bibinfo{journal}{Phys. Rev. Lett.} \textbf{\bibinfo{volume}{93}},
  \bibinfo{pages}{166403} (\bibinfo{year}{2004}),
  \eprint{arXiv:cond-mat/0412001}.

\bibitem[{\citenamefont{Blumenstein et~al.}(2011)\citenamefont{Blumenstein,
  Schafer, Mietke, Meyer, Dollinger, Lochner, Cui, Patthey, Matzdorf, and
  Claessen}}]{blumenstein2011}
\bibinfo{author}{\bibfnamefont{C.}~\bibnamefont{Blumenstein}},
  \bibinfo{author}{\bibfnamefont{J.}~\bibnamefont{Schafer}},
  \bibinfo{author}{\bibfnamefont{S.}~\bibnamefont{Mietke}},
  \bibinfo{author}{\bibfnamefont{S.}~\bibnamefont{Meyer}},
  \bibinfo{author}{\bibfnamefont{A.}~\bibnamefont{Dollinger}},
  \bibinfo{author}{\bibfnamefont{M.}~\bibnamefont{Lochner}},
  \bibinfo{author}{\bibfnamefont{X.~Y.} \bibnamefont{Cui}},
  \bibinfo{author}{\bibfnamefont{L.}~\bibnamefont{Patthey}},
  \bibinfo{author}{\bibfnamefont{R.}~\bibnamefont{Matzdorf}}, \bibnamefont{and}
  \bibinfo{author}{\bibfnamefont{R.}~\bibnamefont{Claessen}},
  \bibinfo{journal}{Nat. Phys.} \textbf{\bibinfo{volume}{7}},
  \bibinfo{pages}{776} (\bibinfo{year}{2011}),
  \urlprefix\url{http://dx.doi.org/10.1038/nphys2051}.

\bibitem[{\citenamefont{Paredes et~al.}(2004)\citenamefont{Paredes, Widera,
  Murg, Mandel, Folling, Cirac, Shlyapnikov, Hansch, and
  Bloch}}]{paredes_toks_experiment}
\bibinfo{author}{\bibfnamefont{B.}~\bibnamefont{Paredes}},
  \bibinfo{author}{\bibfnamefont{A.}~\bibnamefont{Widera}},
  \bibinfo{author}{\bibfnamefont{V.}~\bibnamefont{Murg}},
  \bibinfo{author}{\bibfnamefont{O.}~\bibnamefont{Mandel}},
  \bibinfo{author}{\bibfnamefont{S.}~\bibnamefont{Folling}},
  \bibinfo{author}{\bibfnamefont{I.}~\bibnamefont{Cirac}},
  \bibinfo{author}{\bibfnamefont{G.}~\bibnamefont{Shlyapnikov}},
  \bibinfo{author}{\bibfnamefont{T.}~\bibnamefont{Hansch}}, \bibnamefont{and}
  \bibinfo{author}{\bibfnamefont{I.}~\bibnamefont{Bloch}},
  \bibinfo{journal}{Nature (London)} \textbf{\bibinfo{volume}{429}},
  \bibinfo{pages}{277} (\bibinfo{year}{2004}).

\bibitem[{\citenamefont{Kinoshita et~al.}(2004)\citenamefont{Kinoshita, Wenger,
  and Weiss}}]{kinoshita_tonks_experiment}
\bibinfo{author}{\bibfnamefont{T.}~\bibnamefont{Kinoshita}},
  \bibinfo{author}{\bibfnamefont{T.}~\bibnamefont{Wenger}}, \bibnamefont{and}
  \bibinfo{author}{\bibfnamefont{D.}~\bibnamefont{Weiss}},
  \bibinfo{journal}{Science} \textbf{\bibinfo{volume}{305}},
  \bibinfo{pages}{5687} (\bibinfo{year}{2004}).

\bibitem[{\citenamefont{{Liao} et~al.}(2010)\citenamefont{{Liao}, {Rittner},
  {Paprotta}, {Li}, {Partridge}, {Hulet}, {Baur}, and {Mueller}}}]{liao2010}
\bibinfo{author}{\bibfnamefont{Y.}~\bibnamefont{{Liao}}},
  \bibinfo{author}{\bibfnamefont{A.~S.~C.} \bibnamefont{{Rittner}}},
  \bibinfo{author}{\bibfnamefont{T.}~\bibnamefont{{Paprotta}}},
  \bibinfo{author}{\bibfnamefont{W.}~\bibnamefont{{Li}}},
  \bibinfo{author}{\bibfnamefont{G.~B.} \bibnamefont{{Partridge}}},
  \bibinfo{author}{\bibfnamefont{R.~G.} \bibnamefont{{Hulet}}},
  \bibinfo{author}{\bibfnamefont{S.~K.} \bibnamefont{{Baur}}},
  \bibnamefont{and} \bibinfo{author}{\bibfnamefont{E.~J.}
  \bibnamefont{{Mueller}}}, \bibinfo{journal}{Nature (London)}
  \textbf{\bibinfo{volume}{467}}, \bibinfo{pages}{567} (\bibinfo{year}{2010}),
  \eprint{0912.0092}.

\bibitem[{\citenamefont{Cazalilla et~al.}(2011)\citenamefont{Cazalilla, Citro,
  Giamarch, Orignac, and Rigol}}]{cazalilla_review_11}
\bibinfo{author}{\bibfnamefont{M.}~\bibnamefont{Cazalilla}},
  \bibinfo{author}{\bibfnamefont{R.}~\bibnamefont{Citro}},
  \bibinfo{author}{\bibfnamefont{T.}~\bibnamefont{Giamarch}},
  \bibinfo{author}{\bibfnamefont{E.}~\bibnamefont{Orignac}}, \bibnamefont{and}
  \bibinfo{author}{\bibfnamefont{M.}~\bibnamefont{Rigol}}
  (\bibinfo{year}{2011}), \bibinfo{note}{cond-mat/1101.5337}.

\bibitem[{\citenamefont{Voit}(1995)}]{voit_bosonization_revue}
\bibinfo{author}{\bibfnamefont{J.}~\bibnamefont{Voit}}, \bibinfo{journal}{Rep.
  Prog. Phys.} \textbf{\bibinfo{volume}{58}}, \bibinfo{pages}{977}
  (\bibinfo{year}{1995}).

\bibitem[{\citenamefont{Giamarchi}(2004)}]{giamarchi_book_1d}
\bibinfo{author}{\bibfnamefont{T.}~\bibnamefont{Giamarchi}},
  \emph{\bibinfo{title}{Quantum Physics in One Dimension}}
  (\bibinfo{publisher}{Oxford University Press}, \bibinfo{address}{Oxford},
  \bibinfo{year}{2004}).

\bibitem[{\citenamefont{Javanainen}(1995)}]{Javanainen95}
\bibinfo{author}{\bibfnamefont{J.}~\bibnamefont{Javanainen}},
  \bibinfo{journal}{Phys. Rev. Lett.} \textbf{\bibinfo{volume}{75}},
  \bibinfo{pages}{1927} (\bibinfo{year}{1995}),
  \urlprefix\url{http://link.aps.org/doi/10.1103/PhysRevLett.75.1927}.

\bibitem[{\citenamefont{Graham and Walls}(1996)}]{Graham96}
\bibinfo{author}{\bibfnamefont{R.}~\bibnamefont{Graham}} \bibnamefont{and}
  \bibinfo{author}{\bibfnamefont{D.}~\bibnamefont{Walls}},
  \bibinfo{journal}{Phys. Rev. Lett.} \textbf{\bibinfo{volume}{76}},
  \bibinfo{pages}{1774} (\bibinfo{year}{1996}),
  \urlprefix\url{http://link.aps.org/doi/10.1103/PhysRevLett.76.1774}.

\bibitem[{\citenamefont{{Stewart} et~al.}(1998)\citenamefont{{Stewart},
  Gaebler, and { Jin}}}]{one}
\bibinfo{author}{\bibfnamefont{J.~T.} \bibnamefont{{Stewart}}},
  \bibinfo{author}{\bibfnamefont{J.}~\bibnamefont{Gaebler}}, \bibnamefont{and}
  \bibinfo{author}{\bibfnamefont{D.~S.} \bibnamefont{{ Jin}}},
  \bibinfo{journal}{Nature} \textbf{\bibinfo{volume}{454}},
  \bibinfo{pages}{744} (\bibinfo{year}{1998}).

\bibitem[{\citenamefont{Kozuma et~al.}(1999)\citenamefont{Kozuma, Deng, Hagley,
  Wen, Lutwak, Helmerson, Rolston, and Phillips}}]{two}
\bibinfo{author}{\bibfnamefont{M.}~\bibnamefont{Kozuma}},
  \bibinfo{author}{\bibfnamefont{L.}~\bibnamefont{Deng}},
  \bibinfo{author}{\bibfnamefont{E.~W.} \bibnamefont{Hagley}},
  \bibinfo{author}{\bibfnamefont{J.}~\bibnamefont{Wen}},
  \bibinfo{author}{\bibfnamefont{R.}~\bibnamefont{Lutwak}},
  \bibinfo{author}{\bibfnamefont{K.}~\bibnamefont{Helmerson}},
  \bibinfo{author}{\bibfnamefont{S.~L.} \bibnamefont{Rolston}},
  \bibnamefont{and} \bibinfo{author}{\bibfnamefont{W.~D.}
  \bibnamefont{Phillips}}, \bibinfo{journal}{Phys. Rev. Lett.}
  \textbf{\bibinfo{volume}{82}}, \bibinfo{pages}{871} (\bibinfo{year}{1999}).

\bibitem[{\citenamefont{Stenger et~al.}(1999)\citenamefont{Stenger, Inouye,
  Chikkatur, Stamper-Kurn, Pritchard, and Ketterle}}]{three}
\bibinfo{author}{\bibfnamefont{J.}~\bibnamefont{Stenger}},
  \bibinfo{author}{\bibfnamefont{S.}~\bibnamefont{Inouye}},
  \bibinfo{author}{\bibfnamefont{A.~P.} \bibnamefont{Chikkatur}},
  \bibinfo{author}{\bibfnamefont{D.~M.} \bibnamefont{Stamper-Kurn}},
  \bibinfo{author}{\bibfnamefont{D.~E.} \bibnamefont{Pritchard}},
  \bibnamefont{and} \bibinfo{author}{\bibfnamefont{W.}~\bibnamefont{Ketterle}},
  \bibinfo{journal}{Phys. Rev. Lett.} \textbf{\bibinfo{volume}{82}},
  \bibinfo{pages}{4569} (\bibinfo{year}{1999}).

\bibitem[{\citenamefont{Stamper-Kurn et~al.}(1999)\citenamefont{Stamper-Kurn,
  Chikkatur, G\"orlitz, Inouye, Gupta, Pritchard, and Ketterle}}]{four}
\bibinfo{author}{\bibfnamefont{D.~M.} \bibnamefont{Stamper-Kurn}},
  \bibinfo{author}{\bibfnamefont{A.~P.} \bibnamefont{Chikkatur}},
  \bibinfo{author}{\bibfnamefont{A.}~\bibnamefont{G\"orlitz}},
  \bibinfo{author}{\bibfnamefont{S.}~\bibnamefont{Inouye}},
  \bibinfo{author}{\bibfnamefont{S.}~\bibnamefont{Gupta}},
  \bibinfo{author}{\bibfnamefont{D.~E.} \bibnamefont{Pritchard}},
  \bibnamefont{and} \bibinfo{author}{\bibfnamefont{W.}~\bibnamefont{Ketterle}},
  \bibinfo{journal}{Phys. Rev. Lett.} \textbf{\bibinfo{volume}{83}},
  \bibinfo{pages}{2876} (\bibinfo{year}{1999}).

\bibitem[{\citenamefont{Ozeri et~al.}(2005)\citenamefont{Ozeri, Katz,
  Steinhauer, and Davidson}}]{five}
\bibinfo{author}{\bibfnamefont{R.}~\bibnamefont{Ozeri}},
  \bibinfo{author}{\bibfnamefont{N.}~\bibnamefont{Katz}},
  \bibinfo{author}{\bibfnamefont{J.}~\bibnamefont{Steinhauer}},
  \bibnamefont{and} \bibinfo{author}{\bibfnamefont{N.}~\bibnamefont{Davidson}},
  \bibinfo{journal}{Rev. Mod. Phys.} \textbf{\bibinfo{volume}{77}},
  \bibinfo{pages}{187} (\bibinfo{year}{2005}).

\bibitem[{\citenamefont{Papp et~al.}(2008)\citenamefont{Papp, Pino, Wild,
  Ronen, Wieman, Jin, and Cornell}}]{six}
\bibinfo{author}{\bibfnamefont{S.~B.} \bibnamefont{Papp}},
  \bibinfo{author}{\bibfnamefont{J.~M.} \bibnamefont{Pino}},
  \bibinfo{author}{\bibfnamefont{R.~J.} \bibnamefont{Wild}},
  \bibinfo{author}{\bibfnamefont{S.}~\bibnamefont{Ronen}},
  \bibinfo{author}{\bibfnamefont{C.~E.} \bibnamefont{Wieman}},
  \bibinfo{author}{\bibfnamefont{D.~S.} \bibnamefont{Jin}}, \bibnamefont{and}
  \bibinfo{author}{\bibfnamefont{E.~A.} \bibnamefont{Cornell}},
  \bibinfo{journal}{Phys. Rev. Lett.} \textbf{\bibinfo{volume}{101}},
  \bibinfo{pages}{135301} (\bibinfo{year}{2008}).

\bibitem[{\citenamefont{Veeravalli et~al.}(2008)\citenamefont{Veeravalli,
  Kuhnle, Dyke, and Vale}}]{seven}
\bibinfo{author}{\bibfnamefont{G.}~\bibnamefont{Veeravalli}},
  \bibinfo{author}{\bibfnamefont{E.}~\bibnamefont{Kuhnle}},
  \bibinfo{author}{\bibfnamefont{P.}~\bibnamefont{Dyke}}, \bibnamefont{and}
  \bibinfo{author}{\bibfnamefont{C.~J.} \bibnamefont{Vale}},
  \bibinfo{journal}{Phys. Rev. Lett.} \textbf{\bibinfo{volume}{101}},
  \bibinfo{pages}{250403} (\bibinfo{year}{2008}).

\bibitem[{\citenamefont{{Du} et~al.}(2010)\citenamefont{{Du}, {Wan},
  {Yesilada}, {Ryu}, {Heinzen}, {Liang}, and {Wu}}}]{eleven}
\bibinfo{author}{\bibfnamefont{X.}~\bibnamefont{{Du}}},
  \bibinfo{author}{\bibfnamefont{S.}~\bibnamefont{{Wan}}},
  \bibinfo{author}{\bibfnamefont{E.}~\bibnamefont{{Yesilada}}},
  \bibinfo{author}{\bibfnamefont{C.}~\bibnamefont{{Ryu}}},
  \bibinfo{author}{\bibfnamefont{D.~J.} \bibnamefont{{Heinzen}}},
  \bibinfo{author}{\bibfnamefont{Z.}~\bibnamefont{{Liang}}}, \bibnamefont{and}
  \bibinfo{author}{\bibfnamefont{B.}~\bibnamefont{{Wu}}}, \bibinfo{journal}{New
  Journal of Physics} \textbf{\bibinfo{volume}{12}}, \bibinfo{pages}{083025}
  (\bibinfo{year}{2010}), \eprint{0704.2623}.

\bibitem[{\citenamefont{Fabbri et~al.}(2009)\citenamefont{Fabbri, Cl\'ement,
  Fallani, Fort, Modugno, van~der Stam, and Inguscio}}]{twelve}
\bibinfo{author}{\bibfnamefont{N.}~\bibnamefont{Fabbri}},
  \bibinfo{author}{\bibfnamefont{D.}~\bibnamefont{Cl\'ement}},
  \bibinfo{author}{\bibfnamefont{L.}~\bibnamefont{Fallani}},
  \bibinfo{author}{\bibfnamefont{C.}~\bibnamefont{Fort}},
  \bibinfo{author}{\bibfnamefont{M.}~\bibnamefont{Modugno}},
  \bibinfo{author}{\bibfnamefont{K.~M.~R.} \bibnamefont{van~der Stam}},
  \bibnamefont{and} \bibinfo{author}{\bibfnamefont{M.}~\bibnamefont{Inguscio}},
  \bibinfo{journal}{Phys. Rev. A} \textbf{\bibinfo{volume}{79}},
  \bibinfo{pages}{043623} (\bibinfo{year}{2009}).

\bibitem[{\citenamefont{Cl\'ement et~al.}(2009)\citenamefont{Cl\'ement, Fabbri,
  Fallani, Fort, and Inguscio}}]{thirth}
\bibinfo{author}{\bibfnamefont{D.}~\bibnamefont{Cl\'ement}},
  \bibinfo{author}{\bibfnamefont{N.}~\bibnamefont{Fabbri}},
  \bibinfo{author}{\bibfnamefont{L.}~\bibnamefont{Fallani}},
  \bibinfo{author}{\bibfnamefont{C.}~\bibnamefont{Fort}}, \bibnamefont{and}
  \bibinfo{author}{\bibfnamefont{M.}~\bibnamefont{Inguscio}},
  \bibinfo{journal}{Phys. Rev. Lett.} \textbf{\bibinfo{volume}{102}},
  \bibinfo{pages}{155301} (\bibinfo{year}{2009}).

\bibitem[{\citenamefont{{Ernst} et~al.}(2010)\citenamefont{{Ernst},
  {G{\"o}tze}, {Krauser}, {Pyka}, {L{\"u}hmann}, {Pfannkuche}, and
  {Sengstock}}}]{ernst_bragg}
\bibinfo{author}{\bibfnamefont{P.~T.} \bibnamefont{{Ernst}}},
  \bibinfo{author}{\bibfnamefont{S.}~\bibnamefont{{G{\"o}tze}}},
  \bibinfo{author}{\bibfnamefont{J.~S.} \bibnamefont{{Krauser}}},
  \bibinfo{author}{\bibfnamefont{K.}~\bibnamefont{{Pyka}}},
  \bibinfo{author}{\bibfnamefont{D.-S.} \bibnamefont{{L{\"u}hmann}}},
  \bibinfo{author}{\bibfnamefont{D.}~\bibnamefont{{Pfannkuche}}},
  \bibnamefont{and}
  \bibinfo{author}{\bibfnamefont{K.}~\bibnamefont{{Sengstock}}},
  \bibinfo{journal}{Nature Physics} \textbf{\bibinfo{volume}{6}},
  \bibinfo{pages}{56} (\bibinfo{year}{2010}).

\bibitem[{\citenamefont{Menotti et~al.}(2003)\citenamefont{Menotti, Kr\"amer,
  Pitaevskii, and Stringari}}]{fourth}
\bibinfo{author}{\bibfnamefont{C.}~\bibnamefont{Menotti}},
  \bibinfo{author}{\bibfnamefont{M.}~\bibnamefont{Kr\"amer}},
  \bibinfo{author}{\bibfnamefont{L.}~\bibnamefont{Pitaevskii}},
  \bibnamefont{and}
  \bibinfo{author}{\bibfnamefont{S.}~\bibnamefont{Stringari}},
  \bibinfo{journal}{Phys. Rev. A} \textbf{\bibinfo{volume}{67}},
  \bibinfo{pages}{053609} (\bibinfo{year}{2003}).

\bibitem[{\citenamefont{{Roth} and {Burnett}}(2004)}]{fiveth}
\bibinfo{author}{\bibfnamefont{R.}~\bibnamefont{{Roth}}} \bibnamefont{and}
  \bibinfo{author}{\bibfnamefont{K.}~\bibnamefont{{Burnett}}},
  \bibinfo{journal}{J. Phys. B} \textbf{\bibinfo{volume}{37}},
  \bibinfo{pages}{3893} (\bibinfo{year}{2004}).

\bibitem[{\citenamefont{Hofstetter et~al.}(2002)\citenamefont{Hofstetter,
  Cirac, Zoller, Demler, and Lukin}}]{sixth}
\bibinfo{author}{\bibfnamefont{W.}~\bibnamefont{Hofstetter}},
  \bibinfo{author}{\bibfnamefont{J.~I.} \bibnamefont{Cirac}},
  \bibinfo{author}{\bibfnamefont{P.}~\bibnamefont{Zoller}},
  \bibinfo{author}{\bibfnamefont{E.}~\bibnamefont{Demler}}, \bibnamefont{and}
  \bibinfo{author}{\bibfnamefont{M.~D.} \bibnamefont{Lukin}},
  \bibinfo{journal}{Phys. Rev. Lett.} \textbf{\bibinfo{volume}{89}},
  \bibinfo{pages}{220407} (\bibinfo{year}{2002}).

\bibitem[{\citenamefont{van Oosten et~al.}(2005)\citenamefont{van Oosten,
  Dickerscheid, Farid, van~der Straten, and Stoof}}]{seventh}
\bibinfo{author}{\bibfnamefont{D.}~\bibnamefont{van Oosten}},
  \bibinfo{author}{\bibfnamefont{D.~B.~M.} \bibnamefont{Dickerscheid}},
  \bibinfo{author}{\bibfnamefont{B.}~\bibnamefont{Farid}},
  \bibinfo{author}{\bibfnamefont{P.}~\bibnamefont{van~der Straten}},
  \bibnamefont{and} \bibinfo{author}{\bibfnamefont{H.~T.~C.}
  \bibnamefont{Stoof}}, \bibinfo{journal}{Phys. Rev. A}
  \textbf{\bibinfo{volume}{71}}, \bibinfo{pages}{021601}
  (\bibinfo{year}{2005}).

\bibitem[{\citenamefont{Rey et~al.}(2005)\citenamefont{Rey, Blakie, Pupillo,
  Williams, and Clark}}]{eightth}
\bibinfo{author}{\bibfnamefont{A.~M.} \bibnamefont{Rey}},
  \bibinfo{author}{\bibfnamefont{P.~B.} \bibnamefont{Blakie}},
  \bibinfo{author}{\bibfnamefont{G.}~\bibnamefont{Pupillo}},
  \bibinfo{author}{\bibfnamefont{C.~J.} \bibnamefont{Williams}},
  \bibnamefont{and} \bibinfo{author}{\bibfnamefont{C.~W.} \bibnamefont{Clark}},
  \bibinfo{journal}{Phys. Rev. A} \textbf{\bibinfo{volume}{72}},
  \bibinfo{pages}{023407} (\bibinfo{year}{2005}).

\bibitem[{\citenamefont{Pupillo et~al.}(2006)\citenamefont{Pupillo, Rey, and
  Batrouni}}]{nineth}
\bibinfo{author}{\bibfnamefont{G.}~\bibnamefont{Pupillo}},
  \bibinfo{author}{\bibfnamefont{A.~M.} \bibnamefont{Rey}}, \bibnamefont{and}
  \bibinfo{author}{\bibfnamefont{G.~G.} \bibnamefont{Batrouni}},
  \bibinfo{journal}{Phys. Rev. A} \textbf{\bibinfo{volume}{74}},
  \bibinfo{pages}{013601} (\bibinfo{year}{2006}).

\bibitem[{\citenamefont{Mathey et~al.}(2009)\citenamefont{Mathey, Danshita, and
  Clark}}]{twentyth}
\bibinfo{author}{\bibfnamefont{L.}~\bibnamefont{Mathey}},
  \bibinfo{author}{\bibfnamefont{I.}~\bibnamefont{Danshita}}, \bibnamefont{and}
  \bibinfo{author}{\bibfnamefont{C.~W.} \bibnamefont{Clark}},
  \bibinfo{journal}{Phys. Rev. A} \textbf{\bibinfo{volume}{79}},
  \bibinfo{pages}{011602} (\bibinfo{year}{2009}).

\bibitem[{\citenamefont{Titvinidze et~al.}(2008)\citenamefont{Titvinidze,
  Snoek, and Hofstetter}}]{twentyoneth}
\bibinfo{author}{\bibfnamefont{I.}~\bibnamefont{Titvinidze}},
  \bibinfo{author}{\bibfnamefont{M.}~\bibnamefont{Snoek}}, \bibnamefont{and}
  \bibinfo{author}{\bibfnamefont{W.}~\bibnamefont{Hofstetter}},
  \bibinfo{journal}{Phys. Rev. Lett.} \textbf{\bibinfo{volume}{100}},
  \bibinfo{pages}{100401} (\bibinfo{year}{2008}).

\bibitem[{\citenamefont{Golovach et~al.}(2009)\citenamefont{Golovach, Minguzzi,
  and Glazman}}]{golovach_2009}
\bibinfo{author}{\bibfnamefont{V.~N.} \bibnamefont{Golovach}},
  \bibinfo{author}{\bibfnamefont{A.}~\bibnamefont{Minguzzi}}, \bibnamefont{and}
  \bibinfo{author}{\bibfnamefont{L.~I.} \bibnamefont{Glazman}},
  \bibinfo{journal}{Phys. Rev. A} \textbf{\bibinfo{volume}{80}},
  \bibinfo{pages}{043611} (\bibinfo{year}{2009}),
  \urlprefix\url{http://link.aps.org/doi/10.1103/PhysRevA.80.043611}.

\bibitem[{\citenamefont{{Chotia} et~al.}(2011)\citenamefont{{Chotia},
  {Neyenhuis}, {Moses}, {Yan}, {Covey}, {Foss-Feig}, {Rey}, {Jin}, and
  {Ye}}}]{Chotia_2011}
\bibinfo{author}{\bibfnamefont{A.}~\bibnamefont{{Chotia}}},
  \bibinfo{author}{\bibfnamefont{B.}~\bibnamefont{{Neyenhuis}}},
  \bibinfo{author}{\bibfnamefont{S.~A.} \bibnamefont{{Moses}}},
  \bibinfo{author}{\bibfnamefont{B.}~\bibnamefont{{Yan}}},
  \bibinfo{author}{\bibfnamefont{J.~P.} \bibnamefont{{Covey}}},
  \bibinfo{author}{\bibfnamefont{M.}~\bibnamefont{{Foss-Feig}}},
  \bibinfo{author}{\bibfnamefont{A.~M.} \bibnamefont{{Rey}}},
  \bibinfo{author}{\bibfnamefont{D.~S.} \bibnamefont{{Jin}}}, \bibnamefont{and}
  \bibinfo{author}{\bibfnamefont{J.}~\bibnamefont{{Ye}}},
  \emph{\bibinfo{title}{{Long-lived dipolar molecules and Feshbach molecules in
  a 3D optical lattice}}} (\bibinfo{year}{2011}), \eprint{arXiv:1110.4420}.

\bibitem[{\citenamefont{Lahaye et~al.}(2007)\citenamefont{Lahaye, Koch,
  Fr\"ohlich, Fattori, Metz, Griesmaier, Giovanazzi, and Pfau}}]{pfau_a0}
\bibinfo{author}{\bibfnamefont{T.}~\bibnamefont{Lahaye}},
  \bibinfo{author}{\bibfnamefont{T.}~\bibnamefont{Koch}},
  \bibinfo{author}{\bibfnamefont{B.}~\bibnamefont{Fr\"ohlich}},
  \bibinfo{author}{\bibfnamefont{M.}~\bibnamefont{Fattori}},
  \bibinfo{author}{\bibfnamefont{J.}~\bibnamefont{Metz}},
  \bibinfo{author}{\bibfnamefont{A.}~\bibnamefont{Griesmaier}},
  \bibinfo{author}{\bibfnamefont{S.}~\bibnamefont{Giovanazzi}},
  \bibnamefont{and} \bibinfo{author}{\bibfnamefont{T.}~\bibnamefont{Pfau}},
  \bibinfo{journal}{Nature (London)} \textbf{\bibinfo{volume}{448}},
  \bibinfo{pages}{672} (\bibinfo{year}{2007}).

\bibitem[{\citenamefont{Lu et~al.}(2011)\citenamefont{Lu, Burdick, Youn, and
  Lev}}]{lu2011}
\bibinfo{author}{\bibfnamefont{M.}~\bibnamefont{Lu}},
  \bibinfo{author}{\bibfnamefont{N.~Q.} \bibnamefont{Burdick}},
  \bibinfo{author}{\bibfnamefont{S.~H.} \bibnamefont{Youn}}, \bibnamefont{and}
  \bibinfo{author}{\bibfnamefont{B.~L.} \bibnamefont{Lev}},
  \bibinfo{journal}{Phys. Rev. Lett.} \textbf{\bibinfo{volume}{107}},
  \bibinfo{pages}{190401} (\bibinfo{year}{2011}),
  \urlprefix\url{http://link.aps.org/doi/10.1103/PhysRevLett.107.190401}.

\bibitem[{\citenamefont{Citro et~al.}(2008)\citenamefont{Citro, {De Palo},
  Orignac, Pedri, and Chiofalo}}]{ours}
\bibinfo{author}{\bibfnamefont{R.}~\bibnamefont{Citro}},
  \bibinfo{author}{\bibfnamefont{S.}~\bibnamefont{{De Palo}}},
  \bibinfo{author}{\bibfnamefont{E.}~\bibnamefont{Orignac}},
  \bibinfo{author}{\bibfnamefont{P.}~\bibnamefont{Pedri}}, \bibnamefont{and}
  \bibinfo{author}{\bibfnamefont{M.-L.} \bibnamefont{Chiofalo}},
  \bibinfo{journal}{New J. Phys.} \textbf{\bibinfo{volume}{10}},
  \bibinfo{pages}{045011} (\bibinfo{year}{2008}).

\bibitem[{\citenamefont{Sokol}(1987)}]{Sokol_BEC}
\bibinfo{author}{\bibfnamefont{P.~E.} \bibnamefont{Sokol}},
  \bibinfo{journal}{Can. J. Phys.} \textbf{\bibinfo{volume}{65}},
  \bibinfo{pages}{1393} (\bibinfo{year}{1987}),
  \urlprefix\url{http://www.nrcresearchpress.com/doi/abs/10.1139/p87-219}.

\bibitem{brunello2001} A. Brunello, 
F. Dalfovo, L. Pitaevskii, S. Stringari, and F. Zambelli, Phys. Rev. A
{\bf 64}, 063614 (2001).  

\bibitem[{\citenamefont{Maslov and Stone}(1995)}]{maslov_pure_wire}
\bibinfo{author}{\bibfnamefont{D.~L.}~\bibnamefont{Maslov}} \bibnamefont{and}
  \bibinfo{author}{\bibfnamefont{M.}~\bibnamefont{Stone}},
  \bibinfo{journal}{Phys. Rev. B} \textbf{\bibinfo{volume}{52}},
  \bibinfo{pages}{R5539} (\bibinfo{year}{1995}).

\bibitem[{\citenamefont{Safi and Schulz}(1995)}]{safi_pure_wire}
\bibinfo{author}{\bibfnamefont{I.}~\bibnamefont{Safi}} \bibnamefont{and}
  \bibinfo{author}{\bibfnamefont{H.~J.} \bibnamefont{Schulz}},
  \bibinfo{journal}{Phys. Rev. B} \textbf{\bibinfo{volume}{52}},
  \bibinfo{pages}{R17040} (\bibinfo{year}{1995}).

\bibitem[{\citenamefont{Fazio et~al.}(1998)\citenamefont{Fazio, Hekking, and
  Khmelnitskii}}]{fazio_thermal_1d}
\bibinfo{author}{\bibfnamefont{R.}~\bibnamefont{Fazio}},
  \bibinfo{author}{\bibfnamefont{F.~W.~J.} \bibnamefont{Hekking}},
  \bibnamefont{and} \bibinfo{author}{\bibfnamefont{D.~E.}
  \bibnamefont{Khmelnitskii}}, \bibinfo{journal}{Phys. Rev. Lett.}
  \textbf{\bibinfo{volume}{80}}, \bibinfo{pages}{5611} (\bibinfo{year}{1998}).

\bibitem[{\citenamefont{Petrov et~al.}(2000)\citenamefont{Petrov, Walraven, and
  Shlyapnikov}}]{petrov_trapped_bosons}
\bibinfo{author}{\bibfnamefont{D.~S.} \bibnamefont{Petrov}},
  \bibinfo{author}{\bibfnamefont{J.}~\bibnamefont{Walraven}}, \bibnamefont{and}
  \bibinfo{author}{\bibfnamefont{G.~V.} \bibnamefont{Shlyapnikov}},
  \bibinfo{journal}{Phys. Rev. Lett.} \textbf{\bibinfo{volume}{85}},
  \bibinfo{pages}{3745} (\bibinfo{year}{2000}).

\bibitem[{\citenamefont{Petrov et~al.}(2004)\citenamefont{Petrov, Gangardt, and
  Shlyapnikov}}]{petrov04_bec_review}
\bibinfo{author}{\bibfnamefont{D.}~\bibnamefont{Petrov}},
  \bibinfo{author}{\bibfnamefont{D.}~\bibnamefont{Gangardt}}, \bibnamefont{and}
  \bibinfo{author}{\bibfnamefont{G.}~\bibnamefont{Shlyapnikov}},
  \bibinfo{journal}{J. de Phys. IV} \textbf{\bibinfo{volume}{116}},
  \bibinfo{pages}{3} (\bibinfo{year}{2004}).

\bibitem[{\citenamefont{Lieb and Liniger}(1963)}]{lieb_bosons_1D}
\bibinfo{author}{\bibfnamefont{E.~H.} \bibnamefont{Lieb}} \bibnamefont{and}
  \bibinfo{author}{\bibfnamefont{W.}~\bibnamefont{Liniger}},
  \bibinfo{journal}{Phys. Rev.} \textbf{\bibinfo{volume}{130}},
  \bibinfo{pages}{1605} (\bibinfo{year}{1963}).

\bibitem[{\citenamefont{Lieb}(1963)}]{lieb_excit}
\bibinfo{author}{\bibfnamefont{E.~H.} \bibnamefont{Lieb}},
  \bibinfo{journal}{Phys. Rev.} \textbf{\bibinfo{volume}{130}},
  \bibinfo{pages}{1616} (\bibinfo{year}{1963}).

\bibitem[{\citenamefont{Girardeau}(1960)}]{girardeau_bosons1d}
\bibinfo{author}{\bibfnamefont{M.}~\bibnamefont{Girardeau}},
  \bibinfo{journal}{J. Math. Phys.} \textbf{\bibinfo{volume}{1}},
  \bibinfo{pages}{516} (\bibinfo{year}{1960}).

\bibitem[{\citenamefont{Menotti and Stringari}(2002)}]{menotti02_bose_hydro1d}
\bibinfo{author}{\bibfnamefont{C.}~\bibnamefont{Menotti}} \bibnamefont{and}
  \bibinfo{author}{\bibfnamefont{S.}~\bibnamefont{Stringari}},
  \bibinfo{journal}{Phys. Rev. A} \textbf{\bibinfo{volume}{66}},
  \bibinfo{pages}{043610} (\bibinfo{year}{2002}),
  \urlprefix\url{http://www.citebase.org/abstract?id=oai:arXiv.org:cond-mat/02%
01158}.

\bibitem[{\citenamefont{Abramowitz and
  Stegun}(1972)}]{abramowitz_math_functions}
\bibinfo{author}{\bibfnamefont{M.}~\bibnamefont{Abramowitz}} \bibnamefont{and}
  \bibinfo{author}{\bibfnamefont{I.}~\bibnamefont{Stegun}},
  \emph{\bibinfo{title}{Handbook of mathematical functions}}
  (\bibinfo{publisher}{Dover}, \bibinfo{address}{New York},
  \bibinfo{year}{1972}).

\bibitem[{\citenamefont{Gradshteyn and Ryzhik}(1980)}]{gradshteyn80_tables}
\bibinfo{author}{\bibfnamefont{A.}~\bibnamefont{Gradshteyn}} \bibnamefont{and}
  \bibinfo{author}{\bibfnamefont{R.}~\bibnamefont{Ryzhik}},
  \emph{\bibinfo{title}{Tables of integrals series and products}}
  (\bibinfo{publisher}{Academic Press}, \bibinfo{address}{New-York},
  \bibinfo{year}{1980}).

\bibitem[{\citenamefont{{Vignolo} et~al.}(2001)\citenamefont{{Vignolo},
  {Minguzzi}, and {Tosi}}}]{vignolo2001}
\bibinfo{author}{\bibfnamefont{P.}~\bibnamefont{{Vignolo}}},
  \bibinfo{author}{\bibfnamefont{A.}~\bibnamefont{{Minguzzi}}},
  \bibnamefont{and} \bibinfo{author}{\bibfnamefont{M.~P.}
  \bibnamefont{{Tosi}}}, \bibinfo{journal}{Phys. Rev. A}
  \textbf{\bibinfo{volume}{64}}, \bibinfo{pages}{023421}
  (\bibinfo{year}{2001}), \eprint{arXiv:cond-mat/0102340}.

\bibitem[{\citenamefont{{Vignale} et~al.}(1997)\citenamefont{{Vignale},
  {Ullrich}, and {Conti}}}]{Vignale97}
\bibinfo{author}{\bibfnamefont{G.}~\bibnamefont{{Vignale}}},
  \bibinfo{author}{\bibfnamefont{C.~A.} \bibnamefont{{Ullrich}}},
  \bibnamefont{and} \bibinfo{author}{\bibfnamefont{S.}~\bibnamefont{{Conti}}},
  \bibinfo{journal}{Physical Review Letters} \textbf{\bibinfo{volume}{79}},
  \bibinfo{pages}{4878} (\bibinfo{year}{1997}),
  \eprint{arXiv:cond-mat/9706306}.

\bibitem[{\citenamefont{{Chiofalo} et~al.}(1998)\citenamefont{{Chiofalo},
  {Minguzzi}, and {Tosi}}}]{Chiofalo98}
\bibinfo{author}{\bibfnamefont{M.~L.} \bibnamefont{{Chiofalo}}},
  \bibinfo{author}{\bibfnamefont{A.}~\bibnamefont{{Minguzzi}}},
  \bibnamefont{and} \bibinfo{author}{\bibfnamefont{M.~P.}
  \bibnamefont{{Tosi}}}, \bibinfo{journal}{Physica B}
  \textbf{\bibinfo{volume}{254}}, \bibinfo{pages}{188} (\bibinfo{year}{1998}),
  \eprint{arXiv:cond-mat/9807064}.

\end{thebibliography}
\end{document}